\documentclass[11pt,reqno]{article}

\usepackage{amsmath}
\usepackage{amsfonts}
\usepackage{amssymb}
\usepackage{amsxtra}
\usepackage{graphicx}
\usepackage{caption}
\usepackage{ushort}
\usepackage{color}
\usepackage{hyperref}
\hypersetup{linktocpage=true}

\makeatletter
\@addtoreset{equation}{section}
\makeatother

\def\nn{\nonumber} \def\bd{\begin{document}} \def\ed{\end{document}}
\def\ds{\documentstyle} \let\fr=\frac \let\bl=\bigl \let\br=\bigr
\let\Br=\Bigr \let\Bl=\Bigl
\let\bm=\bibitem
\let\na=\nabla
\let\pa=\partial \let\ov=\overline

\newcommand{\be}{\begin{equation}}
\newcommand{\ee}{\end{equation}}
\newcommand{\bse}{\begin{subequations}}
\newcommand{\ese}{\end{subequations}}
\newcommand{\bea}{\begin{eqnarray}}
\newcommand{\eea}{\end{eqnarray}}
\newcommand{\ba}{\begin{array}}
\newcommand{\ea}{\end{array}}

\def\ft#1#2{{\textstyle{{\scriptstyle #1}\over {\scriptstyle #2}}}}
\def\fft#1#2{{#1 \over #2}}\def\del{\partial}
\def\vp{\varphi}
\def\sst#1{{\scriptscriptstyle #1}}
\def\st#1{{\scriptstyle #1}}
\def\oneone{\rlap 1\mkern4mu{\rm l}}
\def\td{\tilde}
\def\wtd{\widetilde}
\def\ie{{\it i.e.\ }}
\def\iec{{\it i.e.,\ }}
\def\eg{{\it e.g.\ }}
\def\egc{{\it e.g.,\ }}
\def\dalemb#1#2{{\vbox{\hrule height .#2pt
        \hbox{\vrule width.#2pt height#1pt \kern#1pt
                \vrule width.#2pt}
        \hrule height.#2pt}}}
\def\smsquare{\mathord{\dalemb{6.8}{7}\hbox{\hskip1pt}}}
\newcommand{\ho}[1]{$\, ^{#1}$}
\newcommand{\hoch}[1]{$\, ^{#1}$}
\newcommand{\ra}{\rightarrow}
\newcommand{\lra}{\longrightarrow}
\newcommand{\Lra}{\Leftrightarrow}
\newcommand{\ap}{\alpha^\prime}
\newcommand{\bp}{\tilde \beta^\prime}
\newcommand{\tr}{{\rm tr} }
\newcommand{\Tr}{{\rm Tr} }
\def\0{{\sst{(0)}}}
\def\1{{\sst{(1)}}}
\def\2{{\sst{(2)}}}
\def\3{{\sst{(3)}}}
\def\4{{\sst{(4)}}}
\def\5{{\sst{(5)}}}
\def\6{{\sst{(6)}}}
\def\7{{\sst{(7)}}}
\def\8{{\sst{(8)}}}
\def\9{{\sst{(9)}}}
\def\ten{{\sst{(10)}}}
\def\n{{\sst{(n)}}}
\def\cA{{{\cal A}}}
\def\cF{{{\cal F}}}
\def\tV{\widetilde V}
\def\tW{\widetilde W}
\def\tH{\widetilde H}
\def\tE{\widetilde E}
\def\tF{\widetilde F}
\def\tA{\widetilde A}
\def\im{{{\rm i}}}
\def\tY{{{\wtd Y}}}
\def\ep{{\epsilon}}
\def\vep{{\varepsilon}}
\def\bD{{{\bar D}}}
\def\alp{{{\a'}^3}}
\def\bD{{{\bar D}}}
\def\R{{{\mathbb R}}}
\def\C{{{\mathbb C}}}
\def\E{{{\mathbb E}}}
\def\H{{{\mathbb H}}}
\def\CP{{{\mathbb C}{\mathbb P}}}
\def\RP{{{\mathbb R}{\mathbb P}}}
\def\Z{{{\mathbb Z}}}
\def\bA{{{\mathbb A}}}
\def\bB{{{\mathbb B}}}
\def\bC{{{\mathbb C}}}
\def\bR{{{\mathbb R}}}
\def\bD{{{\mathbb D}}}
\def\bE{{{\mathbb E}}}
\def\bZ{{{\mathbb Z}}}
\def\Re{{{\frak{Re}}}}
\def\Im{{{\frak{Im}}}}
\def\cosec{{\,\hbox{cosec}\,}}
\def\Gm{{\Gamma_{\!\! -}}}
\def\Gp{{\Gamma_{\!\! +}}}

\def\cosech{{\hbox{cosech}}}
\def\sech{{\hbox{sech}}}

\newcommand{\tamphys}{\it Center for Theoretical Physics,
Texas A\&M University, College Station, TX 77843, USA}

\newcommand{\mitchell}{\it George P. \& Cynthia W.
Mitchell Institute for Fundamental Physics,\\
Texas A\&M University, College Station, TX 77843-4242, USA}
\newcommand{\umich}{\it Michigan Center for Theoretical Physics,
University of Michigan\\ Ann Arbor, MI 48109, USA}
\newcommand{\upenn}{\it Department of Physics and Astronomy,
University of Pennsylvania, Philadelphia,  PA 19104, USA}
\newcommand{\SISSA}{\it  SISSA-ISAS and INFN, Sezione di Trieste\\
Via Beirut 2-4, I-34013, Trieste, Italy}

\newcommand{\newton}{\it Isaac Newton Institute for Mathematical Sciences,\\
20 Clarkson Road,  University of Cambridge,
Cambridge CB3 0EH, UK}

\newcommand{\ihp}{\it Institut Henri Poincar\'e\\
  11 rue Pierre et Marie Curie, F 75231 Paris Cedex 05}

\newcommand{\damtp}{\it DAMTP, Centre for Mathematical Sciences,
 Cambridge University\\  Wilberforce Road, Cambridge CB3 OWA, UK}
\newcommand{\itp}{\it Institute for Theoretical Physics, University of
California\\ Santa Barbara, CA 93106, USA}

\newcommand{\imperial}{\it The Blackett Laboratory, Imperial College London\\
Prince Consort Road, London SW7 2AZ}

\newcommand{\beijingnormal}{\it Department of Physics, Beijing Normal University\\
Beijing 100875, China}

\newcommand{\auth}{
B. Crampton\,\footnote{email: benedict.crampton@gmail.com}\hoch{\star},
C.N. Pope\,\footnote{email: pope@physics.tamu.edu}\hoch{\dagger,\ddagger} 
and K.S. Stelle\,\footnote{email: k.stelle@imperial.ac.uk}\hoch{\star}}

\begin{document}
\setcounter{page}{0}
\thispagestyle{empty}
\begin{flushright}
\hfill{MIFPA-14-24 \ \ \ \ \  
Imperial/TP/14/KSS/02}\\
\end{flushright} 

\begin{center}  

{\Large {\bf Braneworld Localisation in Hyperbolic Spacetime}}   

\vspace{15pt}

\auth

\vspace{7pt}
{\hoch{\star}\imperial}

\vspace{7pt}
{\hoch{\dagger}\mitchell}

\vspace{7pt}
{\hoch{\ddagger}\damtp}

\vspace{30pt}

\underline{ABSTRACT}
\end{center} 

We present a construction employing a type IIA supergravity and 
3-form flux background together with an NS5-brane that localises massless 
gravity near the 5-brane worldvolume. The nonsingular underlying type IIA 
solution is a lift to 10D of the vacuum solution of the 6D Salam-Sezgin model 
and has a hyperbolic ${\cal H}^{(2,2)}\times S^1$ structure in the lifting 
dimensions. A fully back-reacted solution including the NS5-brane is 
constructed by recognising the 10D Salam-Sezgin vacuum solution as 
a ``brane resolved through transgression.'' 
The background hyperbolic structure plays a key r\^ole in generating a mass 
gap in the spectrum of the transverse-space wave operator, which gives rise 
to the localisation of gravity on the 6D NS5-brane worldvolume, or, equally, 
in a further compactification to 4D. Also key to the successful 
localisation of gravity is the specific form of the corresponding transverse 
wavefunction Schr\"odinger problem, which asymptotically involves 
a $V=-1/(4\rho^2)$ potential, where $\rho$ is the transverse-space radius, 
and for which the NS5-brane source gives rise to a specific choice of 
self-adjoint extension for the transverse wave operator. The corresponding 
boundary condition as $\rho\to0$ ensures the masslessness of gravity in 
the effective braneworld theory. Above the mass gap, there is a continuum 
of massive states which give rise to small corrections to Newton's law.

\vfill\leftline{}\vfill

\pagebreak

\tableofcontents
\addtocontents{toc}{\protect\setcounter{tocdepth}{2}}
\newpage 

\section{Introduction: the problem of localising gravity on a brane}\label{sec:intro}

   The problem of how to localise gravity on a submanifold of a higher-dimensional spacetime has been a key concern for cosmological braneworld models since the beginnings of the subject \cite{Rubakov:1983bb}. With compact extra dimensions, this is not a main concern, because there is a natural eigenvalue gap between a zero-mode of the Laplacian for the transverse dimensions and the first excited state, giving a corresponding mass gap in the effective 4D theory spectrum between massless 4D gravitational modes and the lowest lying massive modes. With noncompact extra dimensions, however, the problem that arises in principal is to avoid having a continuum of massive states ranging down all the way to those corresponding to massless 4D gravity.
   
   An approach to the localisation of gravity on a 4D subsurface of an infinite higher-dimensional spacetime was given in \cite{Randall:1999vf, Karch:2000ct}, joining segments of $\hbox{AdS}_5$ with the junction providing a 
delta-function source to the Einstein equations which gives rise to a normalisable bound state in the corresponding effective Schr\"odinger problem. Similar constructions involving excisions of spacetime were made, for example  
in \cite{Brandhuber:1999hb}. A problem with such constructions, however, is to realise the delta-function source as a natural brane construct in string or M-theory. An embedding of the 5D $\Z_2$ symmetric construction of Ref.\ \cite{Randall:1999vf} was given in \cite{Duff:2000az}, but lifting the 5D realisation up to 10D proved to involve a singularity with no clear brane or orbifold 
interpretation, located at the lift of the $\Z_2$ 
reflection point \cite{Lehners:2007xa}.
   
   An analysis of the difficulties of realising lower-dimensional gravity, massless or massive, on a subsurface of an infinite higher-dimensional spacetime was given in Ref.\  \cite{Bachas:2011xa}. For string constructions with asymptotically maximally-symmetric spacetimes (de Sitter, Poincar\'e, or anti-de Sitter), it proves to be difficult to obtain the peak in the warp factor for the 4D subspace that is needed in order to give rise to the localising bound state.
   
   Another theme in the study of supergravity theories which has been somewhat explored but not widely applied is the existence of supergravity models with noncompact gauge symmetries (see, for example, \cite{Hull:1988jw}). Such gaugings may elegantly be obtained using the embedding-tensor formalism \cite{deWit:2005ub}. Models with gauged noncompact group symmetries of this sort manage to have a purely positive-energy spectrum thanks to the nonlinear realisation of the noncompact symmetry on appropriate sets of scalar fields, acting on prefix factors of positive-energy kinetic terms, with linear realisation only on a compact subgroup of the gauged symmetry. One reason that few proposed physical applications of higher-dimensional models with noncompact gauge symmetries have been made, however, is the generally continuous spectrum of eigenstates in the space transverse to the lower-dimensional spacetime. The corresponding continuous spectrum of effective-theory massive states can prevent the effective localisation of lower-dimensional gravity, unless somehow a mass gap can be arranged below the edge of the continuous spectrum.
   
   In this paper, we combine these two developments to provide a construction that localises gravity on a subspace of a background spacetime arising from just such a noncompact gauged supergravity. Instead of a simple patching of slices of the background spacetime, however, our construction employs a natural object in string or M-theory: an NS5-brane. This is one of the fundamental brane objects arising in 11D M-theory \cite{Gueven:1992hh}, and it gives rise, upon ``vertical'' dimensional reduction \cite{Lu:1996mg}, to the NS5-brane of type IIA supergravity. 
   
   The construction is ultimately based upon a 6D model with R-symmetry gauging obtained by Salam and Sezgin in 1984 \cite{Salam:1984cj}. This model has 
the unusual property of having a scalar field with a {\em positive} 
potential, as opposed to the negative or indefinite sign potentials arising in models with gauged compact symmetries. Although the Salam-Sezgin model does not admit a maximally symmetric 6D spacetime solution, the positive scalar potential does allow for a solution combining an $S^2$ subspace with $U(1)$ magnetic monopole flux and with flat 4D Minkowski space.  The link between the Salam-Sezgin model and supergravities related to string theory is given by its embedding into 10D type IIA supergravity by a lift on the noncompact space ${\cal H}^{(2,2)}$ \cite{Cvetic:2003xr}. When viewed as a 7D supergravity theory, the full theory obtained {\it via} ${\cal H}^{(2,2)}$ reduction from 10D has a gauged ${\rm SO}(2,2)$ symmetry, positive-energy kinetic terms for all fields and 9 scalar fields with a positive-definite potential generalising that of the Salam-Sezgin model (which can then be obtained by a consistent truncation of the ${\rm SO}(2,2)$ invariant theory). If desired, the construction can be extended to an 11D embedding by the inclusion of an additional lift on a further spatial $S^1$.
   
   In Section \ref{sec:SSembeddingH22}, we begin by presenting the details of the embedding of the Salam-Sezgin 6D $S^2\times \R^4$ ``vacuum'' solution into a 10D type IIA supergravity solution {\it via} a Kaluza-Klein lift on ${\cal H}^{(2,2)}$. This then sets the stage in Section \ref{sec:boundstates} for an initial analysis of gravitational fluctuations about the Salam-Sezgin background, and for a discussion of normalisable candidate bound states that could localise gravity in a lower-dimensional subspacetime. For spin-two excitations, a simplifying feature of such analysis is that one needs only to study the scalar wave equation in the space transverse to the lower-dimensional spacetime \cite{Csaki:2000fc, Bachas:2011xa}. The only such wavefunction that can  explicitly be given in terms of standard functions turns out to be the zero-eigenvalue eigenfunction $\xi_0$ of the transverse wave operator. Gravitational fluctuations with this transverse wavefunction structure are massless from the point of view of the lower-dimensional physics. However, the wavefunction $\xi_0$ has a logarithmic asymptotic behaviour as one approaches the ``waist'' ($\rho=0$ in the radial coordinate) of the ${\cal H}^{(2,2)}$ space, as distinct from the non-singular structure of the underlying Salam-Sezgin vacuum. This implies the need for a source at $\rho=0$ in the fluctuation wave equation. 
   
   Preserving the eight-supercharge unbroken supersymmetry of the vacuum Salam-Sezgin solution points to a NS5-brane as the relevant inclusion, as analysed in Section \ref{sec:SSNS5}, which then proceeds on to a main result: the fully back-reacted solution generalising the Salam-Sezgin vacuum background by the inclusion of an NS5-brane. The key to this construction is recognition of the nonsingular vacuum Salam-Sezgin solution, when reduced from 10D to 9D by compactification on the NS5-brane ``waist'' coordinate $\psi$, as an instance of a ``brane resolved by transgression'' in the fashion of Ref.\ \cite{Cvetic:2000mh}. 
   
   In Section \ref{sec:sourceinclusion}, the needed NS5-brane worldvolume source action is included in the field equations, yielding, firstly, the relation between the NS5-brane tension $T$ and the integration constant $k$ found for the bulk solution inclusion of the NS5-brane, and secondly, the boundary conditions for transverse fluctuation wavefunctions that are required by the NS5-brane source. This analysis is surprisingly subtle, especially concerning the question of self-adjointness of the transverse wave operator: the corresponding Schr\"odinger problem for the transverse wavefunction involves, asymptotically as $\rho\to0$, a potential $V=-1/(4\rho^2)$, which has represented a continuing puzzle in quantum mechanics since the 1950's \cite{Case:1950an, Landau1960, de Alfaro:1976je, Essin2005}. Analysis of the NS5-brane source action's implications for the asymptotic $\rho\to0$ structure of the transverse wavefunction, with careful regulation of the corresponding source delta function,  selects just one transverse bound state. This is the zero mode $\xi_0$, which happily remains exactly the same as in the preliminary fluctuation analysis about the Salam-Sezgin vacuum solution given in Section \ref{sec:SSembeddingH22}. This yields massless gravity in the lower-dimensional braneworld. Moreover, as one moves away from the $\rho=0$ ``waist'' of the ${\cal H}^{(2,2)}$ space, the Schr\"odinger potential rises to a positive value $(1+k)$, depending on the strength $k$ of the NS5-brane. This gives rise to a $(1+k)g^2$ gap in the lower-dimensional braneworld $(\hbox{mass})^2$ eigenvalues between the massless states and the edge of the continuum massive spectrum, where $g$ is the $(\hbox{length})^{-1}$ dimensional parameter characterising the scale of the ${\cal H}^{(2,2)}$ hyperbolic geometry.
   
   The lower-dimensional effective braneworld gravity arising from this construction is initially six-dimensional, corresponding to the worldvolume dimension of the NS5-brane. Of these six  worldvolume dimensions, the ``waist'' coordinate $\psi$ is naturally compactified. Another worldvolume coordinate, $y$, may be chosen to be compactified on $S^1$, or can be used in an $S^1/\Z_2$ Ho\v{r}ava-Witten \cite{Horava:1996ma} type construction in order to produce a 4D chiral theory, with attendant matter fields as needed to cancel anomaly inflow \cite{Lukas:1999nh, Pugh:2010ii}. The 4D effective gravity is next analysed. Normalisation of the $\xi_0$ bound-state transverse wavefunction is carried out explicitly in Section \ref{sec:Newtonconstant} and a preliminary consideration of corrections to 4D Newtonian gravity arising from the continuum of massive states is given in Section \ref{sec:gravitycorrections}. The paper concludes in Section \ref{sec:conclusion} with a consideration of open problems and the realisation of our construction in string theory.

\section{Salam-Sezgin theory and its embedding in \texorpdfstring{${\cal H}^{(2,2)}$}{H(2,2)}}\label{sec:SSembeddingH22}

    The bosonic sector of the six-dimensional Salam-Sezgin theory is described
by the Lagrangian
\be
\bar{\cal L}_6 = \bar R\, {\bar*\oneone} -
 \ft14 {\bar*d\bar\phi}\wedge d\bar\phi -\ft12 e^{\ft12\bar\phi}\,
{\bar* \bar F_\2}\wedge\bar F_\2 -\ft12 e^{\bar\phi}\, 
 {\bar * \bar H_\3}\wedge \bar H_\3 -8\bar g^2\, 
    e^{-\ft12\bar\phi}\, {\bar*\oneone}\,,
\ee
where $d\bar H_\3= \ft12 \bar F_2\wedge \bar F_\2$ and $\bar F_2=d\bar A_\1$.
(We put a bar on all quantities in the six-dimensional theory.)  It was
shown in \cite{Cvetic:2003xr} that the Salam-Sezgin theory can be
embedded in the ten-dimensional
type I supergravity theory\footnote{It can thus also be embedded into type IIA theory, as we shall do later.} whose bosonic Lagrangian can be taken to be
\be
{\cal L}_{10} = R\, {*\oneone} -\ft12 {*d\phi}\wedge d\phi -
\ft12 e^{-\phi}{*F_3}\wedge F_\3\,,
\ee
via a consistent dimensional reduction on $\R$ times the 
three-dimensional hyperbolic space 
${\cal H}^{(2,2)}$.  This space can be defined as the surface
$X_1^2+X_2^2-X_3^2-X_4^2= 1$ embedded in the Euclidean space $\E^4$, 
with the natural metric inherited as the restriction of $ds_{\rm Euclidean}^2= dX_1^2 +
dX_2^2 + dX_3^2 +dX_4^2$ to this surface.  Its
isometry group is $U(1)\times U(1)$, which is the intersection of 
the ${\rm O}(2,2)$ symmetry of the embedding condition and the ${\rm O}(4)$
symmetry of the Euclidean metric on $\E^4$.  It was shown in 
\cite{Cvetic:2003xr} that the metric on ${\cal H}^{(2,2)}$ can be written
as
\be
ds_3^2 = \cosh2\rho\, d\rho^2 + \cosh^2\rho\, d\alpha^2 +
\sinh^2\rho\, d\beta^2\,,
\ee
where $\rho\ge0$ and $0\le\alpha<2\pi$, $0\le\beta<2\pi$.  It will be more
convenient for our purposes to introduce coordinates $\psi\in[0,4\pi)$ and $\chi\in[0,2\pi)$ in
place of $\alpha$ and $\beta$, defined by
\be
\psi= \alpha +\beta\,,\qquad \chi=\alpha -\beta\,.\label{psichi}
\ee
In terms of these, the embedding of the Salam-Sezgin theory in ten-dimensional
type I supergravity that was constructed in \cite{Cvetic:2003xr} is
given by\newpage
\bea
ds_{10}^2 &=& (\cosh2\rho)^{1/4}\, \Big[e^{-\ft14\bar\phi}\, d\bar s_6^2 +
  e^{\ft14\bar\phi}\, dy^2 + \ft12\bar g^{-2}\, 
   e^{\ft14\bar\phi}\, \Big(d\rho^2 \cr
&&+\ft14[d\psi+ 
   \sech2\rho(d\chi-2\bar g \bar A)]^2 + \ft14(\tanh2\rho)^2\,
  (d\chi-2\bar g \bar A)^2\Big)\Big]\,,\cr
F_\3 &=&\bar H_\3 -\fft{\sinh2\rho}{4\bar g^2 (\cosh2\rho)^2}\, 
d\rho\wedge d\psi\wedge(d\chi-2\bar g\bar A_\1)\cr
&&+\fft1{4\bar g\cosh2\rho}\,
\bar F_2\wedge[d\psi + \cosh2\rho\, (d\chi-2\bar g\bar A_\1)]\,,\cr
e^\phi &=& (\cosh2\rho)^{-1/2}\, e^{-\ft12\bar\phi}\,.\label{tentosix}
\eea

   For the present, our focus will be on the remarkable 
(Minkowski)$_4\times S^2$ vacuum solution of the Salam-Segin theory
\cite{Salam:1984cj}, which is given by
\bea
d\bar s_6^2 &=& dx^\mu dx^\nu\eta_{\mu\nu} + \fft1{8\bar g^2}\, 
 (d\theta^2+\sin^2\theta d\varphi^2)\,,\cr
\bar A_\1&=& -\fft1{2\bar g}\, \cos\theta\, d\varphi\,,\qquad 
\bar H_\3=0\,,\qquad \bar\phi=0\,.\label{SS6vac}
\eea
The lift of this solution to ten dimensions, using \eqref{tentosix}, was
given in \cite{Cvetic:2003xr}.  As noted there, the solution is more elegantly
written in the ten-dimensional string-frame metric $ds^2_{\rm str}$, 
related to the Einstein-frame metric $ds_{10}^2$ by
\be
ds^2_{10\,\rm str}= e^{\ft12\phi}\, ds_{10}^2\,.\label{stringEinsteinrel}
\ee
After making the coordinate transformation \eqref{psichi} the
lifted Salam-Sezgin vacuum is given by \cite{Cvetic:2003xr}\,\footnote{As
in \cite{Cvetic:2003xr}, it is convenient to re-express the gauge coupling
constant $\bar g$ of the Salam-Sezgin theory in terms of a new constant
\bea
g= \sqrt2\, \bar g\,,\label{SS10vac}
\eea
and we shall use $g$ from now on.  (This rescaling was done in order to 
avoid $\sqrt2$ factors in the general reduction ansatz.)}
\bea
ds^2_{10 \,\rm str} &=& dx^\mu dx^\nu\eta_{\mu\nu} + dy^2 + \fft1{4g^2}\, 
[d\psi + \sech2\rho\, (d\chi + \cos\theta\, d\varphi)]^2 +
 \fft1{g^2}\, \sech2\rho\, ds^2_{\rm\sst{EH}}\,,\cr
e^\phi &=& (\sech2\rho)^{1/2}\,,\qquad
 A_\2= \fft1{4 g^2}\, [d\chi + \sech2\rho\, d\psi]\wedge 
    (d\chi + \cos\theta\, d\varphi)\,,\label{liftedsalsez}
\eea
where
\be
ds^2_{\rm\sst{EH}} = \cosh2\rho\, d\rho^2 + 
   \fft{(\sinh2\rho)^2}{4\cosh2\rho}\, (d\chi+\cos\theta\, d\varphi)^2
+ \ft14 \cosh2\rho\, (d\theta^2+\sin^2\theta\, d\varphi^2)\,.\label{ehmet}
\ee
We now observe that if we make the coordinate transformation $\cosh2\rho=r^2$,
the metric $ds^2_{\rm\sst{EH}}$ becomes
\be
ds^2_{\rm\sst{EH}}= \Big(1-\fft1{r^4}\Big)^{-1}\, dr^2
+ \ft14 r^2\, \Big(1-\fft1{r^4}\Big)\, (d\chi+\cos\theta\, d\varphi)^2
+ \ft14 r^2\, (d\theta^2+\sin^2\theta\, d\varphi^2)\,.
\ee
This can be recognised as the Eguchi-Hanson metric, with unit scale
parameter \cite{Eguchi:1978gw}.  Recalling that the $\chi$ coordinate has period $2\pi$,
one sees from \eqref{ehmet} that at large distance the space approaches $\R^4/\Z_2$ \cite{Belinsky:1978ue}.
On the other hand, as $\rho$ goes to zero, $\rho$ and $\chi$ become like plane
polar coordinates in the neighbourhood of the origin, so the space 
approaches $\R^2\times S^2$ there.

   The Salam-Sezgin (Minkowski)$_4\times S^2$ vacuum 
is supersymmetric \cite{Salam:1984cj}
in six dimensions, and hence it lifts to a supersymmetric solution in
ten dimensions.  The general reduction of the fermions was discussed
also in \cite{Cvetic:2003xr}.  For our present purposes, it is useful
just to exhibit the Killing spinors of the lifted Salam-Sezgin
vacuum \eqref{liftedsalsez}.  These are most elegantly expressed in the
string frame also, wherein the ten-dimensional supersymmetry transformation
rules take the form
\be
\delta\psi_M = \nabla_M\,\ep 
    - \ft18 F_{MNP}\, \Gamma^{NP}\,\Gamma_{11}\, \ep\,,
\qquad
\delta\lambda = \Gamma^M\del_M\phi\, \ep - 
    \ft1{12} F_{MNP}\,\Gamma^{MNP}\, \Gamma_{11}\, \ep\,.\label{susytrans}
\ee
We shall give a more detailed discussion of the derivation of the
Killing spinors later, when we consider a modification of the 
ten-dimensional lift of the Salam-Sezgin vacuum in which a singular 
NS5-brane is introduced.  For now we shall just present the result for
the lifted Salam-Sezgin vacuum itself.  We find that 
there exist eight Killing spinors, which are given by
\be
\ep = e^{-\ft12\chi\, \Gamma_{89}}\, \eta\,,\label{kspin}
\ee
where $\eta$ is any {\it constant} spinor satisfying the two projection
conditions
\be
\Gamma_{11}\, \eta= -\eta\,,\qquad
\Gamma_{67}\eta= \Gamma_{89}\, \eta\,.\label{proj}
\ee
Here, the 6, 7, 8 and 9 vielbein indices on the gamma matrices refer to
the four directions in the Eguchi-Hanson transverse space, with
\bea
\hat e^6 &=& \ft12 \sinh 2\rho\, (\cosh2\rho)^{-1/2}\,
(d\chi + \cos\theta\, d\varphi)\,,\qquad
\hat e^7 = (\cosh 2\rho)^{1/2}\, d\rho\,,\cr
\hat e^8 &=& \ft12  (\cosh 2\rho)^{1/2}\, d\theta\,,\qquad
\hat e^9= \ft12 (\cosh 2\rho)^{1/2}\,\sin\theta\, d\varphi\,,
\label{EHvielbein}
\eea
and $ds^2_{\rm\sst{EH}}= \sum_{i=6}^9 \hat e_i^2$.

\section{Bound states and mass gaps}\label{sec:boundstates}

Now consider gravitational fluctuations around the Salam-Sezgin background, considered from a braneworld four-dimensional perspective.
General studies \cite{Csaki:2000fc,Bachas:2011xa} of the fluctuation problem about supergravity backgrounds start with an ansatz replacing the 4D metric 
$\eta_{\mu\nu}$ by $\eta_{\mu\nu}+h_{\mu\nu}(x,z)$ where $x^\mu$ are the 
4D coordinates and $z^n=(y,\psi,\theta,\varphi,\chi,\rho)$ are 
the six ``transverse'' 
coordinates. One notes from the Salam-Sezgin background 
solution \eqref{liftedsalsez} that, of these, the 
five coordinates $(y,\psi,\theta,\varphi,\chi)$ all refer to naturally 
compact directions, while $\rho$ is the non-compact ``radius''. The 
fluctuation problem for $h_{\mu\nu}(x,z)$ can be treated by separation of 
variables. A full expansion of the fluctuations in the ten-dimensional 
theory would involve introducing harmonic eigenfunctions for the dependences 
on the five compact coordinates as well as the non-compact $\rho$ 
coordinate. However, for a study of the lowest-lying fluctuation states, one 
may simplify the problem by availing oneself of a consistent truncation to 
the sector independent of $y,\psi,\theta,\varphi\ \&\,\chi$: this amounts 
to considering only S-wave, \ie singlet, states with respect to the 
corresponding background symmetries (\ie $({\rm U}(1))^3$ for 
$y$, $\psi$ and $\chi$, and ${\rm SO}(3)$ for $\theta$ and $\varphi$). 
The essential remaining dependence is then on the non-compact 
coordinate $\rho$. 

Accordingly, we posit an expansion
\be
h_{\mu\nu}(x,\rho)=\sum_i h_{\mu\nu}^{(\lambda_i)}(x)\xi_{\lambda_i}(\rho) 
+ \int_{\Lambda_{\mbox{\scriptsize edge}}}^\infty d\lambda\,
 h_{\mu\nu}^{(\lambda)}(x)\xi_\lambda(\rho)\,,\label{hexpansion}
\ee
where the $\xi_{\lambda_i}$ are discrete states and the 
$\xi_{\lambda}$ are continuum states for eigenvalues $\lambda$ starting from 
some lower value $\Lambda_{\mbox{\scriptsize edge}}$ at the edge of the 
continuous spectrum. Limiting attention to linearised 4D gravitational 
fluctuations in $h_{\mu\nu}(x)$ about the Salam-Sezgin background, we may 
focus on pure spin-two fluctuations with $\eta^{\mu\nu}h_{\mu\nu}(x)=0$ and 
we may also impose the gauge conditions $\partial^\mu h_{\mu\nu}(x)=0$. 
The analysis of \cite{Bachas:2011xa} then shows that the gravitational 
fluctuations must solve a {\em scalar} wave equation in the full 
ten-dimensional spacetime
\be
\square_{(10)}h_{\mu\nu}(x,z)=0\,,\label{10Dwaveeqn}
\ee
where the 10D wave operator splits up as
\be
\square_{(10)}=H_{\sst{\rm SS}}^{\frac14}(\square_{(4)} + 
g^2 \triangle_{y,\psi,\theta,\varphi,\chi} + 
g^2 \triangle_{\mbox{\scriptsize rad}})\,,
\ee
where
\be
H_{\sst{\rm SS}} =(\cosh 2\rho)^{-1}
\ee
is the Salam-Sezgin warp function, $\square_{(4)}$ is the 4D 
d'Alembertian, $\triangle_{y,\psi,\theta,\varphi,\chi}$ is the Laplacian for 
the five compact directions $(y,\psi,\theta,\varphi,\chi)$ (which will 
have zero eigenvalue for our S-wave treatment) and
\be
\triangle_{\mbox{\scriptsize rad}}=
{\partial^2\over\partial \rho^2}+{2\over\tanh(2\rho)}{
\partial\over\partial\rho}\,.\label{radlaplace}
\ee
Solutions to \eqref{10Dwaveeqn} in ten dimensions will then give rise to 4D solutions with $(\hbox{mass})^2$ values given by the $\triangle_{\mbox{\scriptsize rad}}$ eigenvalue $\lambda$:
\bea
 \triangle_{\mbox{\scriptsize rad}}\xi_\lambda(\rho)&=&
-\lambda\xi_\lambda(\rho)\,,\cr
\square_{(4)}h_{\mu\nu}^\lambda(x)&=&m^2h_{\mu\nu}^\lambda(x)\,,\cr
m^2&=&g^2\lambda\,.\label{eigenvalueproblem}
\eea

\subsection{The Schr\"odinger equation for \texorpdfstring{$\triangle_{\mbox{\scriptsize rad}}$}{Laplace(rad)} eigenfunctions}\label{ssec:schrodingereqn}

One can rewrite the $\triangle_{\mbox{\scriptsize rad}}$ eigenvalue problem as a Schr\"odinger equation by making the rescaling
\be
\Psi_\lambda=\sqrt{\sinh(2\rho)}\xi_\lambda\,,\label{rescaledwavefunction}
\ee
thereby  eliminating the first-derivative term from the eigenfunction 
equation, 
which then takes the Schr\"odinger-equation form
\be
-{d^2\Psi_\lambda\over d\rho^2} + 
V(\rho)\Psi_\lambda=\lambda\Psi_\lambda\,,\label{schrodinger}
\ee
where the potential is
\be
V(\rho)=2-{1\over \tanh^2(2\rho)}\,.\label{SSpotential}
\ee

The Schr\"odinger equation potential \eqref{SSpotential} asymptotes to the 
value 1 for large $\rho$. In this large-$\rho$ limit, the 
Schr\"odinger equation becomes
\be
{d^2\Psi_\lambda\over d\rho^2}+4e^{-4\rho}\Psi_\lambda+
(\lambda-1)\Psi_\lambda=0\,,
\ee
giving scattering-state solutions for $\lambda>1$:
\be
\Psi_\lambda(\rho)\sim \left(A_\lambda e^{i\sqrt{\lambda-1}\rho} + 
B_\lambda e^{-i\sqrt{\lambda-1}\rho} \right)\quad\text{ for large $\rho$}\,,
\ee
while for $\lambda<1$ one can have $L^2$ normalizable candidate bound states. 
Recalling the $\rho$ dependence of 
the measure $\sqrt{-g_{(10)}}\sim(\cosh(2\rho))^{\frac14}\sinh(2\rho)$, 
one finds for large $\rho$ that
\be
\int_{\rho_1\gg1}^\infty \vert\Psi_\lambda(\rho)\vert^2d\rho<\infty
\Rightarrow \Psi_\lambda\sim B_\lambda e^{-\sqrt{1-\lambda}\rho} \quad  
\text{for } \lambda<1\,.\label{fallingonly}
\ee
It follows that for $\lambda<1$, we have candidate bound states of the 
transverse $\triangle_{\mbox{\scriptsize rad}}$ system. We shall need to 
discover how to fix the value of $\lambda$. Clearly of particular 
interest will be 
the value zero, which corresponds to massless gravity in four dimensions. 

Before analysing the transverse bound-state spectrum, we need to consider the 
norm that is to be used in considering the normalizability of the transverse 
wavefunction $\xi(\rho)$.  In fact, for the rescaled wavefunction 
\eqref{rescaledwavefunction}, the trivial $L^2$ norm is the correct one. 
One may see this by starting from the 10D type IIA action with 
Einstein-frame 
integration over $\int d^{10}x \sqrt{g_{\ten\,\rm ein}}$. 
With the $\eta_{\mu\nu}\rightarrow\eta_{\mu\nu}+h_{\mu\nu}(x,\rho)$ ansatz 
for the 4D gravitational fluctuations, one obtains the correct norm measure 
by collecting the $\rho$-dependent terms multiplying the 4D gravitational 
kinetic terms. In the quadratic terms of the 4D effective action, up to a 
constant factor of $\frac1{16g^5}$, one has $|\xi(\rho)|^2$ from the 
two $h_{\mu\nu}$ fields, multiplied by 
$H_{\sst{\rm SS}}\sinh 2\rho\cosh2 \rho\sqrt{g_{\2}}$ and then 
integrated over the six dimensions transverse to the 4D space, 
where $\sqrt{g_{\2}}$ is the standard 2-sphere metric density.\footnote{Note 
that this agrees fully with the expression 
$\int d^6y\sqrt{[\hat g]}e^{2A}\xi^2$ given in \cite{Bachas:2011xa} for the 
norm, in which $\sqrt{[\hat g]}$ is the density in the transverse six 
dimensions and $e^{2A}$ is the warp factor of the 4D subspace.}
 The $\rho$-dependent terms combine to give simply $\sinh 2\rho$. 
Consequently, 
after the wavefunction rescaling in \eqref{rescaledwavefunction}, 
the correct norm for the transverse $\rho$-dependent wavefunction is simply
\be
||\Psi||^2 = \int d\rho |\Psi|^2\label{psinorm}\,.
\ee

\subsection{The zero-mass candidate bound state}

The general behaviour of candidate $\Psi_\lambda$ eigenfunctions cannot be 
given in terms of standard functions, but for $\lambda=0$ the Schr\"odinger 
equation happily can be solved exactly, giving the normalised result
\bea
\Psi_0(\rho)&=&\frac{2\sqrt{3}}{\pi}\sqrt{\sinh(2\rho)}\xi_0(\rho)\,,
\label{scaledzeromode}\\
\xi_0(\rho)&=&\log(\tanh\rho)\,.\label{zeromode}
\eea
Sketching the zero-mode wavefunction \eqref{scaledzeromode} and the potential \eqref{SSpotential} together, we have the picture shown in Figure \ref{zeromode+potential}:
\begin{figure}[ht]
\centering
\captionsetup{format=hang}
\includegraphics[scale=.8]{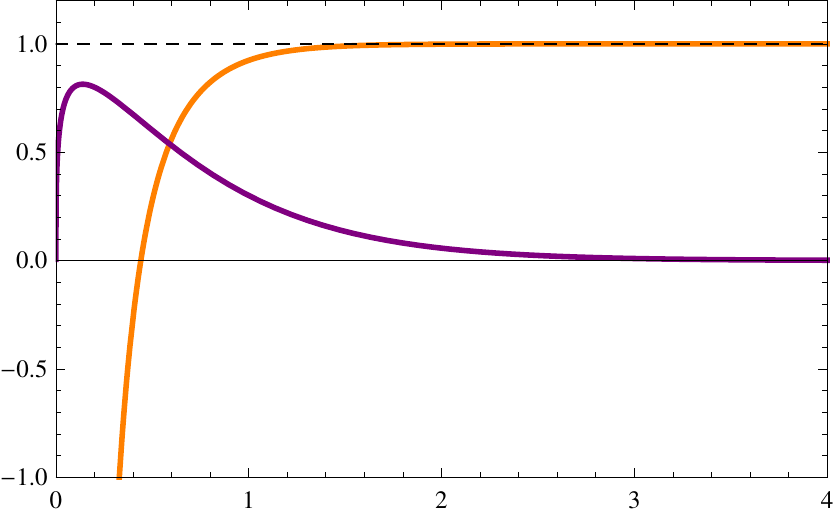}
\caption{The normalizable zero-mode $\Psi_0$ and the ${\cal H}^{(2,2)}$ Schr\"odinger equation potential,\newline\null \ limiting to the value 1 as $\rho\to\infty$.}
\label{zeromode+potential}
\end{figure}

The $(y,\psi,\theta,\varphi,\chi)$ coordinates correspond to a 
compact $T^2\times S^2\times S^1$ space on which one may make a 
standard Kaluza-Klein dimensional reduction. Note that $\chi$ is a 
coordinate corresponding to collapsing curves as one takes the limit 
$\rho\rightarrow0$: the $\rho,\chi$ submanifold simply tends to $\R^2$ in 
this limit; we will come back to this point in the next section. All of the 
other compact coordinates correspond to non-collapsing curves and there is 
no subtlety in restricting attention to fields independent of 
$y$, $\psi$, $\theta$ and $\varphi$ on $T^2\times S^2$. Provided one has 
reason to specify the normalizable $\lambda=0$ wavefunction as the 
remaining part of the $h_{\mu\nu}(x,\rho)$ field dependence on the 
coordinates transverse to the 4D subspace, this will concentrate the 
gravitational fluctuations in the region closely surrounding the 4D 
subspace and will give rise to massless 4D effective gravity. It remains 
now to justify why this zero mode is in fact the correct transverse 
wavefunction.

\section{Salam-Sezgin background with an NS5-brane inclusion}\label{sec:SSNS5}

In the preceding section, we found an attractive zero-mode candidate for the 
gravitational fluctuation wavefunction in the space transverse to our 
4D spacetime. There are, however, two linked aspects of this zero-mode 
wavefunction that require us to expand our consideration of background 
type IIA supergravity solutions in which the 4D gravity-localising subspace 
may be embedded. Although the Salam-Sezgin background \eqref{liftedsalsez} 
is itself a completely smooth solution of type IIA supergravity, the $\xi_0$ 
zero-mode \eqref{zeromode} diverges in the limit as $\rho\rightarrow0$. 
Moreover, as we shall see in detail later, this transverse wavefunction does 
not, strictly speaking, yield a true solution of type IIA supergravity: 
it requires a source at $\rho=0$ (just as the $1/r$ potential requires a 
source in the 3D Laplace equation). This situation is not in itself any 
more disturbing than the need, strictly speaking, for a source for the 
M2 brane \cite{Duff:1990xz}, or for essentially any of the classic string 
or M-theory brane solutions. But the question before us here is: 
a source for what? A number of hints can be found in the Salam-Sezgin 
background solution \eqref{liftedsalsez} and in the logarithmic character 
of the $\xi_0$ zero-mode itself.

The $\log\rho$ behaviour of $\xi_0(\rho)$ as $\rho\rightarrow0$ is a clear 
hint that this wavefunction belongs to a two-dimensional transverse 
space. The natural coordinate to accompany $\rho$ in this transverse 
2-space is $\chi$, which together with $\rho$ comprises polar coordinates
on  $\R^2$, 
in the limit $\rho\rightarrow0$, as mentioned above. If we are looking to 
modify the Salam-Sezgin solution by the inclusion of some kind of brane, 
the natural situation would be to have a flat subspace of the 10D 
solution as the worldvolume. Within the Salam-Sezgin background 
solution \eqref{liftedsalsez}, in addition to the flat 4D coordinates 
$x^\mu$, the $T^2$ coordinates $y$ and $\psi$ are a natural pair of 
coordinates of further flat directions. So the suggestion is to 
consider 
$(x^\mu,y,\psi)$ as world-volume coordinates, with 
$(\rho,\chi,\theta, \varphi)$ as coordinates transverse to a brane 
inclusion. This implies that one should look for a 5-brane inclusion, 
with $(x^\mu,y,\psi)$ as the worldvolume coordinates. Of the 
transverse 
coordinates, we clearly want to focus on solutions depending only on 
$\rho$, and so we will not be exciting modes depending on 
$\theta$, $\varphi$ or 
$\chi$. As we have seen, however, $\chi$, together with $\rho$, form polar 
coordinates on $\R^2$ near $\rho=0$, and although we will not be 
considering functional dependence on $\chi$, care will be needed in 
treating it, since it is part of a suspected operative two-dimensional 
transverse space. The coordinates $(\theta$ and $\varphi$ parameterise 
an $S^2$ in the 
transverse space, on which we will be considering only S-wave, 
\ie $S^2$-independent, solutions. The hints from the structure of 
the 10D Salam-Sezgin solution \eqref{liftedsalsez} therefore point towards the 
inclusion of a 5-brane smeared over the transverse $S^2$ directions, 
thus leaving $\rho$ and $\chi$ as the coordinates of the 
operative two-dimensional transverse space, in which a wave function 
logarithmic in $\rho$ would be natural.

The hints of 5-brane structure in the Salam-Sezgin solution were noted 
already in Reference \cite{Cvetic:2003xr}, where the $\rho\to\infty$ 
asymptotic structure of the Salam-Sezgin background was identified with 
an NS5-brane geometry in ten dimensions, with two of the worldvolume 
coordinates (here $y$ and $\psi$) wrapped around a $T^2$ torus. This 
structure thus makes use of both the ``diagonal'' and ``vertical'' 
dimensional reduction arrangements outlined in Reference \cite{Lu:1996mg}.

A further confirmation that an NS5-brane is the right kind of 
brane 
inclusion to consider comes from supersymmetry. We saw 
in (\ref{kspin}) and (\ref{proj}) that the Salam-Sezgin background has 
eight unbroken supersymmetries. Supersymmetry preservation for a probe 
NS5-brane on this background follows from the requirement of 
$\kappa$-symmetry invariance. For a probe NS5-brane with worldvolume 
directions $(0,1,2,3,4,5)$ in the notation of Section 
\ref{sec:SSembeddingH22}, the corresponding requirement 
is \cite{Skenderis:2002vf}
\be
\Gamma_{012345}\eta=\eta\,.\label{kappacondition}
\ee
The appropriateness of an NS5-brane inclusion can then be seen by 
rewriting the second equation in \eqref{proj} as
\be
\Gamma_{6789}\eta=-\eta\label{6789cond}\,.
\ee
The first equation in \eqref{proj} can be rewritten as
\be
\Gamma_{012345}\Gamma_{6789}\eta=-\eta\,, \label{regamma11cond}
\ee
and so \eqref{6789cond} and \eqref{regamma11cond} together
imply \eqref{kappacondition} already from the Salam-Sezgin supersymmetry 
conditions, resulting in no further diminution of unbroken supersymmetry 
arising from the inclusion of an NS5-brane: the eight unbroken 
Salam-Sezgin supersymmetries persist upon inclusion of an NS5-brane.

These considerations based on the structure of the Salam-Sezgin background 
geometry and probe-brane supersymmetry preservation indicate that the
inclusion of an NS5-brane in the Salam-Sezgin geometry is the appropriate 
way to create an initially static background about which 4D massless 
gravitational fluctuations with a normalizable transverse wavefunction 
can exist. To complete the construction, however, we will need a fully 
back-reacted geometry including an NS5-brane. Constructing this solution 
is now our main task.

\subsection{Lifted Salam-Sezgin vacuum and brane 
resolution by transgression}

   In order to show how an NS5-brane can be introduced, it is helpful
first to perform a dimensional reduction to 9D of the ten-dimensional IIA theory, and of the lifted Salam-Sezgin vacuum, on a circle.  Specifically, we shall 
reduce on the $\psi$ coordinate in (\ref{tentosix}) and (\ref{liftedsalsez}),
or, more precisely, on the rescaled dimensionful coordinate
\be
w \equiv \fft1{2g}\, \psi.
\ee
It is convenient to use the string frames in ten and in nine dimensions,
with the reduction ansatz
\be
ds_{10\,\rm str}^2 = d\tilde s_{9\,\rm str}^2 + e^{\sqrt2\, \tilde\phi_2}\, 
   (dw + \tilde\cA_\1)^2\,,\qquad
A_\2= \tilde A_\2 + \tilde A_\1\wedge dw\,.\label{tentonine}
\ee
(We put tildes on all nine-dimensional fields.)  
The dimensionally-reduced 
theory in nine dimensions, written now in the Einstein frame 
$d\tilde s_{9\,\rm ein}^2=
e^{-2\phi/\sqrt7+ \sqrt2\, \tilde\phi_2/7}\, d\tilde s_{9\,\rm str}^2$, 
is described
by the Lagrangian
\be
\tilde{\cal L}_9 =\sqrt{-\tilde g}\, 
\Big( \tilde R-\ft12(\del\phi_1)^2 -\ft12 (\del\tilde \phi_2)^2
   -\ft1{12}\, e^{\ft{4}{\sqrt{14}}\, \phi_1}\, \tilde F_\3^2 -
    \ft14 e^{\ft{2}{\sqrt{14}}\, \phi_1}\,( e^{-\sqrt2 \tilde \phi_2}\, 
\tilde F_\2^2 +
   e^{\sqrt2\, \tilde \phi_2}\, \tilde \cF_\2^2)\Big)\,,
\ee
where $\phi_1=-\sqrt{\fft87}\, \phi +\fft1{\sqrt7}\, \tilde\phi_2$, and the
nine-dimensional field strengths are given by
\be
\tilde F_\3= d\tilde A_\2 - d\tilde A_\1\wedge \tilde\cA_\1\,,\qquad
\tilde F_\2=d\tilde A_\1\,,\qquad
\tilde \cF_\2=d\tilde\cA_\1\,.\label{d9fields}
\ee

   The nine-dimensional reduction of the lifted Salam-Sezgin vacuum is
given by
\bse\label{transgression1}
\bea
d\tilde s_{9\,\rm str}^2 &=& dx^\mu dx_\mu + dy^2 + \fft1{g^2}\sech2\rho\,
ds_{\rm\sst{EH}}^2\,, \label{transgression1a} \\
\tilde A_\2 &=& \fft1{4g^2}\, \cos\theta\, d\chi\wedge d\varphi\,,\qquad
\tilde A_\1 = -\fft1{2g}\, \sech2\rho\, (d\chi + \cos\theta\, d\varphi)\,,\label{transgression1b}\\
\tilde\cA_\1 &=&\fft1{2g}\, \sech2\rho\, (d\chi + \cos\theta\, d\varphi)\,,
\qquad e^{\sqrt{\ft72}\, \phi_1} = \cosh2\rho\,, \qquad
  \tilde \phi_2=0\,.\label{transgression1c}
\eea
\ese
The 2-form field strengths are therefore given by
\be
\tilde F_2=-\tilde\cF_\2= \fft{-2}{g(\cosh2\rho)^2}\, 
  (\hat e^6\wedge \hat e^7-\hat e^8\wedge \hat e^9)\,,
\label{FcF}
\ee
where $\hat e^i=g(\cosh2\rho)^{\fft12}\tilde e^i$ is the vielbein for the Eguchi-Hanson metric as
defined in \eqref{EHvielbein}.  The 2-form in (\ref{FcF}) can be recognised
as being the normalizable anti-self-dual harmonic 2-form in Eguchi-Hanson geometry.
This was used in \cite{Cvetic:2000mh} to construct so-called ``branes
resolved through transgression,'' and in fact the solution 
(\ref{transgression1}) is precisely an example of this kind.  Applied
to our nine-dimensional case, the procedure described in \cite{Cvetic:2000mh}
allows us to construct resolved 4-brane solutions, with $(x^\mu,y)$ 
being the worldvolume coordinates, since $\tilde F_\3$ obeys
the Bianchi identity $d\tilde F_\3= -\tilde F_\2\wedge \tilde\cF_\2$ 
(see (\ref{d9fields})).
Provided $ds_{\rm\sst{EH}}^2$ is Ricci-flat (which it is here, 
for Eguchi-Hanson space),
and that $\tilde F_\2$ and $\tilde \cF_\2$ are (anti)self-dual in the 
$ds_{\rm\sst{EH}}^2$ metric,
then by making a standard 4-brane ansatz, which is
\be
e^{\sqrt{\ft72}\, \phi_1}\, {\tilde*\tilde F_\3} = d\tilde A_\5\,,\qquad
   \tilde A_\5= H^{-1}\, d^4x \,dy\,,\qquad e^{\sqrt{\ft72}\phi_1} = H^{-1}
\ee
where dualisation is done with respect to the 9D metric \eqref{transgression1a},
we get a solution provided $H$ satisfies
\be
\triangle_{\rm\sst{EH}}\, H = 
   \frac{g^2}2  \tilde F^{ij}\, \tilde \cF_{ij}\,,\label{4dal}
\ee
where the radial part of $\triangle_{\rm\sst{EH}}$ is related to the radial scalar Laplacian \eqref{radlaplace} in the
transverse metric by 
\be
\triangle_{\rm\sst{EH}}=(\cosh2\rho)^{-1}\triangle_{\mbox{\scriptsize rad}}\label{EH-rad_rel}
\ee 
and the $i,j$ indices in \eqref{4dal}
need to be taken in the $\hat e^i$ Eguchi-Hanson 4D basis \eqref{EHvielbein}. For the transverse S-wave solutions 
considered here (\ie for solutions without excitations in the 
$\psi,\theta,\varphi,\chi$ or $y$ variables), we will henceforth just 
write $\triangle_{\rm\sst{EH}}$.

   Plugging (\ref{FcF}) into (\ref{4dal}), we obtain the 
equation 
\be
\triangle_{\sst{\rm EH}}\, H  = -\fft{8}{\cosh^4 2\rho}\,.
\ee
The Salam-Sezgin vacuum itself corresponds to the ''fully-resolved'' solution
\be
H_{\sst{\rm SS}}= \sech 2\rho\,.\label{HSS}
\ee  
The most general solution of the form $H=H(\rho)$ is given, however, by allowing for an additional homogeneous term $\tilde H$ solving $\triangle_{\rm\sst{EH}}\, \tilde H=0$:
\bea
H &=& H_{\sst{\rm SS}} + \tilde H\label{totalH}\\
\tilde H &=& c_1 + c_2\, \log\tanh\rho \,,\label{Hsol}\label{tildeH}
\eea
where $c_1$ and $c_2$ are arbitrary constants. The singularity at $\rho=0$ that
arises when $c_2$ is non-zero corresponds to having an actual 4-brane
solution that will require an appropriate delta-function source, as we shall see in Section \ref{sec:sourceinclusion}.  It will
be necessary to take $c_2$ to be negative in order to obtain a 
well-behaved positive-tension brane solution.  It will also turn out 
that normalizability requirements for the zero-mode fluctuations around
the brane solution imply that we should take $c_1=0$.  Thus from now
on we shall take 
\be
H= \sech 2\rho - k \log\tanh\rho\,,\label{Hsol2}
\ee
where $k$ is a positive constant.

   Finally, we lift the nine-dimensional 4-brane solution back to ten
dimensions using (\ref{tentonine}).  This gives the fully back-reacted metric including the NS5-brane (again in the string frame)
\bea
ds_{10\,\rm str }^2 &=& dx^\mu dx^\nu\eta_{\mu\nu} + dy^2 + 
  \fft1{4 g^2}\, [d\psi + \sech 2\rho\, (d\chi+\cos\theta\, d\varphi)]^2
 + \fft{H}{g^2}\, ds_{\rm\sst{EH}}^2\,,\cr
e^{\phi} &=& H^{1/2}\,,\qquad 
A_\2 = \fft1{4g^2}\, \Big[ (1+k)\, d\chi + \sech 2\rho\, d\psi\Big]
    \wedge (d\chi+ \cos\theta\, d\varphi)\,,\label{ns5brane}
\eea
with $ds_{\rm\sst{EH}}^2$ being the Eguchi-Hanson metric 
(\ref{ehmet}) and where
the function $H$ is given by (\ref{Hsol2}).  
One may return to the Einstein frame using \eqref{stringEinsteinrel}. 
This yields the Einstein-frame form of the full metric including the NS5-brane:
\be
ds_{10\,\rm ein}^2 = H^{-\ft14} ds_{10\,\rm str }^2\,. \label{ns5einsteinframe}
\ee
This is the exact NS5-brane 
generalisation of the lifted Salam-Sezgin vacuum that will form the
basis for our braneworld analysis in the subsequent sections.  Note that
there is a ``twist'' in the $\psi$ worldvolume direction on the
NS5-brane.\footnote{Such kinds of twisted lifts of resolved brane 
solutions have been constructed previously, in \cite{Lu:2002xa}.}  Although
this twist means that there is not a full six-dimensional Poincar\'e 
symmetry of the worldvolume coordinates $(x^\mu, y,\psi)$ of the NS5-brane,
the only essential symmetry for our purposes is the four-dimensional
Poincar\'e symmetry of the four-dimensional spacetime coordinates $x^\mu$.

   At very small $\rho$ we have
\be
H\sim -k\log\rho\,.\label{Hasymp}
\ee
The fact that $H$ has the characteristic form of a harmonic function in two
dimensions rather than the full four dimensions of the transverse space is
a reflection of the fact that near the origin $\rho=0$ the Eguchi-Hanson
space is of the form $\R^2\times S^2$.

\subsection{Supersymmetry of the NS5-brane}

   The general arguments in \cite{Cvetic:2000mh} show that the 
degree of supersymmetry of a ``brane resolved through transgression'' will be the same for any 
solution $H$ of the equation (\ref{4dal}).  Thus we expect in our case that
the inclusion of the NS5-brane in the lifted Salam-Sezgin vacuum, achieved
by taking the constant $k$ in (\ref{Hsol2}) to be non-zero, will give
a background that has the same number of Killing spinors as we found in
Section \ref{sec:SSembeddingH22} for the lifted Salam-Sezgin vacuum itself.  Here, we present 
some results necessary for constructing the Killing spinors in the
NS5-brane background.

 We shall work in the ten-dimensional string frame metric, and so from 
(\ref{ns5brane}) it is natural to choose the vielbein
\bea
e^\mu &=&dx^\mu\,,\ \ \mu=0,1,2,3\,,\ \  e^4=dy\,,\quad 
 e^5 = \fft1{2g}\,[d\psi + \sech2\rho(d\chi+ \cos\theta d\varphi)]\,,
\nn\\
e^i &=& H^{1/2}\,g^{-1}\,  \hat e^i\,,\ \ i= 6,7,8,9\,,\label{10viel}
\eea
where $\hat e^i$ is the Eguchi-Hanson vielbein defined in
(\ref{EHvielbein}). From (\ref{10viel}) we calculate the torsion-free 
spin connection $\omega_{AB}$, which, encapsulated in the spinor
exterior covariant derivative 
$\nabla\equiv d + \ft14 \omega_{AB}\, \Gamma^{AB}$, turns out to be
\bea
\nabla &=& d + \ft14\hat\omega_{ij}\, \Gamma^{ij} 
  -\ft18 g^2H^{-1}\, \tilde\cF_{ij}\, \Gamma^{ij}\,  e^5 -
 \ft14 gH^{-3/2}\, (\sech2\rho)^{1/2}\, H'\, e^j\, \Gamma_{7j}
\nn\\
&&
+ \ft14 g^2H^{-1}\, \tilde\cF_{ij}\, \Gamma^{5i}\,e^j\,,
\eea
where the indices $i$ and $j$ range over the Eguchi-Hanson directions 6, 7, 8
and 9, and $\hat\omega$ is the torsion-free spin connection in the
Eguchi-Hanson space.
   From (\ref{ns5brane}), the 3-form $F_\3=dA_\2$ is given by
\be
F_\3= -\fft{2g}{H (\cosh2\rho)^2}\, e^5\wedge (e^6\wedge e^7- e^8\wedge e^9)
   -\fft{g H'}{H^{3/2}\, (\cosh2\rho)^{1/2}}\, e^6\wedge e^8\wedge e^9\,.
\ee

   Substituting these expressions into the string-frame supersymmetry
transformation rules (\ref{susytrans}), we find that there exist Killing
spinors $\epsilon_{\st{\rm str}}$ given by
\be
\epsilon_{\st{\rm str}} = e^{-\ft12\chi\, \Gamma_{89}}\, \eta\,,\label{kspin2}
\ee
where $\eta$ is any {\it constant} spinor satisfying the two projection
conditions
\be
\Gamma_{11}\, \eta= -\eta\,,\qquad \Gamma_{67}\, \eta= \Gamma_{89}\, \eta\,.
\label{proj2}
\ee
Thus in the string frame, the Killing spinors are identical to those we 
obtained in Section \ref{sec:SSembeddingH22} for the lifted Salam-Sezgin vacuum. In the Einstein frame \eqref{stringEinsteinrel}, the Killing spinor is given by 
\be
\epsilon_{\st{\rm ein}}=e^{-\ft18\phi}\epsilon_{\st{\rm str}}=
H^{-\ft1{16}}\epsilon_{\st{\rm str}}\,.\label{einsteinKilling}
\ee

\section{Inclusion of the NS-5 brane source}\label{sec:sourceinclusion}

Having developed the fully back-reacted solution \eqref{ns5brane} for an 
NS5-brane on the Salam-Sezgin background, we now need to include in the 
action and field equations the corresponding source. Just as the 
solution $V=q/r$ for the three-dimensional Laplace equation requires 
a source, $\triangle V= -4\pi q\delta^3(r)$, or as the M2 brane requires 
a corresponding 2-brane source \cite{Duff:1990xz}, so here we require 
an appropriate source for the NS5-brane. 

The NS5-brane action \cite{Bandos:2002az} is rather complicated on account 
of the worldvolume self-dual 3-form field strength. However, here all we 
really need are the parts that source the Einstein and 10D dilaton equations. 
For this purpose, it is appropriate to use the Einstein-frame form of the 
fully back-reacted metric \eqref{ns5einsteinframe}. Since we are interested 
in S-wave solutions that have ${\rm SO}(3)$ symmetry in the $S^2$ 
directions, the source needs to be smeared over the $S^2$ directions in the 
transverse space. The $\psi$ and $y$ directions of the solution 
\eqref{ns5brane} are worldvolume directions, so the NS5-brane is seen to 
be wrapped around these cycles. The needed NS5-brane source is 
then
\be
I_s={-T\over\Omega_2}\int d^2\Omega \int d^6\zeta
\left[-\det\Big(\partial_i x^M\partial_j x^Ng_{MN}(x(\zeta))
   \Big)\right]^{\frac12}e^{-\phi/2}\,.\label{ns5source}
\ee

The inclusion of the NS5-brane source is depicted in 
Figure \ref{NS5source}:
\begin{figure}[ht]
\centering
\captionsetup{format=hang}
\includegraphics[scale=.45]{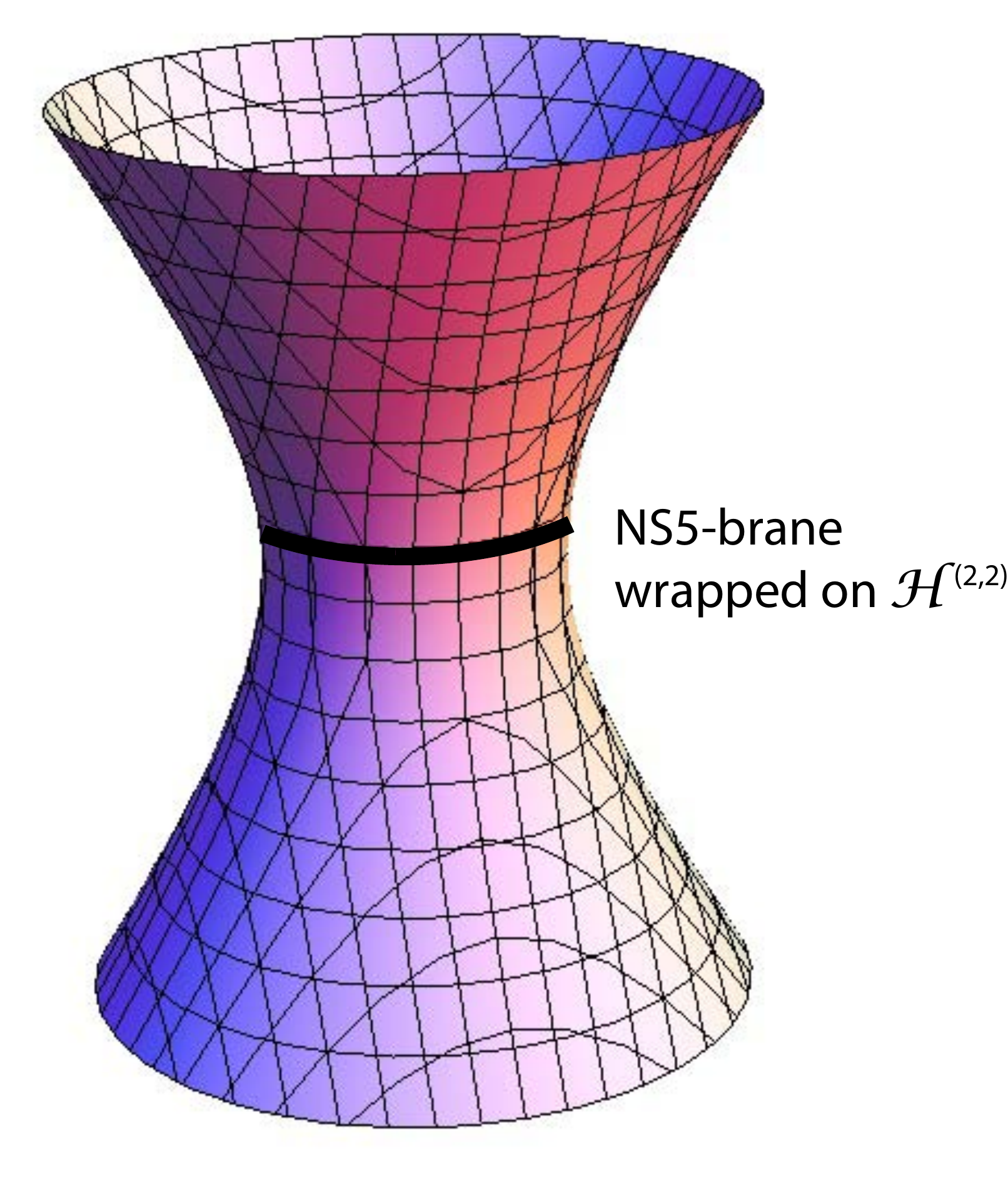}
\caption{NS5-brane source wrapped on the $\psi\in[0,4\pi)$ cycle of ${\cal H}^{(2,2)}$ and smeared in the transverse $S^2$ directions of the bulk solution \eqref{ns5brane}.}
\label{NS5source}
\end{figure}

With the inclusion of this source, the relevant part of the Einstein 
equation for the static Salam-Sezgin + NS5 background, with 4D gravity 
fluctuations, is, after multiplication on both sides by $H^2$,
\be
{1\over 16\pi G_{\sst{10}}}\left(g^2\eta_{\mu\nu}\triangle_{\rm\sst{EH}}
\tilde H -H^2\square_{(4)}  h_{\mu\nu}\xi- g^2H 
h_{\mu\nu}\triangle_{\rm\sst{EH}}\xi\right)= 
-T{g^4\over\sqrt{g_{\rm\sst{EH}}}}(\eta_{\mu\nu} - h_{\mu\nu}\,\xi(\rho))
\delta^2(z)\,,\label{sourcedeqns}
\ee
where $H$ and $\tilde H$ are as in (\ref{totalH}) and (\ref{tildeH}), 
and $G_{\sst{10}}$ is the 10D Newton constant.
Owing to the smearing of the source in the $S^2$ directions, the relevant 
transverse space is reduced to just two dimensions; hence one has just 
the two-dimensional $\delta^2(z)$ delta function, 
for $z^a=(r,\chi)$, on the right-hand side of \eqref{sourcedeqns}.

  For the static background with 
$\tilde H=-k\log\tanh \rho \sim -k\log\rho$, 
Equation \eqref{sourcedeqns} gives
\be
\int\sqrt{g_{\rm\sst{EH}}}\triangle_{\rm\sst{EH}}\tilde H = -16\pi G_{\sst{10}}T g^2\,.\label{staticrel}
\ee
Performing the integral while noting that $\sqrt{g_{\rm\sst{EH}}}=
\frac18\cosh2\rho\sinh2\rho\sin\theta \rightarrow
\frac14\rho\sin\theta $ as $\rho\to0$, and that
\be
d^2z\delta^2(z) = {1\over2\pi}d\rho d\chi\delta(\rho)\label{pointdelta}
\ee 
for $\mbox{SO}(2)$ invariant (\ie $\chi$-independent) functions, and 
using the explicit form of $\tilde H$,
one obtains on the left-hand side 
$\frac\pi2\rho{\partial\over\partial\rho}\tilde H$.  One thus gets the 
following relation between $k$ and $T$:
\be
k=32G_{\sst{10}}Tg^2\,.\label{kTrel}
\ee

\subsection{Fluctuations about the NS5-brane}\label{ssec:fluctuations}

 Having identified the relation \eqref{kTrel} between the 10D bulk solution integration constant $k$ and the NS5-brane source tension $T$, one can confront the analysis of gravitational $h_{\mu\nu}(x,\rho)$ fluctuations around the static $\hbox{Salam-Sezgin}+\hbox{NS5}$ background. From the different factors of $H$ in the sourced equation \eqref{sourcedeqns}, one finds that the sourced eigenvalue problem relevant to 4D gravitational fluctuations with $\hbox{mass}^2=g^2\lambda$ is 
 \be 
{1\over 16\pi G_{\sst{10}}}\left( \triangle_{\rm\sst{EH}}\xi_\lambda(\rho)+\lambda H\xi(\rho) \right) = -T{g^2\over H\sqrt{g_{\rm\sst{EH}}}}\xi_\lambda(\rho)\delta^2(z)\,.\label{sourcedeigenvalue}
 \ee

At this point, one hopes to use the sourced equation \eqref{sourcedeigenvalue} to determine the relevant boundary conditions as $\rho\to0$ for the transverse wavefunction $\xi_\lambda(\rho)$. A standard Frobenius analysis of the eigenfunction problem shows that as $\rho\to 0$, $\xi_\lambda(\rho)$ has the following asymptotic behaviour, both in the original Salam-Sezgin background and also with the NS5-brane inclusion, and for arbitrary eigenvalue $\lambda$:
\be
\xi_\lambda(\rho) \rightarrow a_\lambda+b_\lambda\log\rho\,.\label{xiasymptotics}
\ee

For a transverse bound state, one requires normalizability of the transverse wavefunction $\xi_\lambda(\rho)$ with respect to the full norm, including also the effects of the NS5-brane inclusion via the total $H$ function \eqref{totalH},
\be
||\xi||^2=\frac1{\pi^2}\int d^4z_{\rm\sst{EH}}H\sqrt{g_{\rm\sst{EH}}}|\xi|^2\,.\label{fullnorm}
\ee
Overall normalizability, including all contributions from $\rho\in(0,\infty)$, will determine an overall normalisation constant for $\xi_\lambda$. But we also need to know the asymptotic value for the ratio $a_\lambda/b_\lambda$, which can be parametrised by $\varpi=\arctan({a_\lambda\over b_\lambda})$. Numerical study of the eigenvalue problem shows that there is a one-to-one relationship $\varpi(\lambda)$ between the eigenvalue $\lambda$ and $\varpi$, with $\varpi=0$ corresponding to $\lambda=0$.

Unlike the situation as $\rho\to\infty$, where the requirement of 
normalizability selects the most strongly falling $\xi_\lambda(\rho)$ 
solution with $\lambda<1+k$ for candidate bound state 
wavefunctions,\footnote{Note that $-k\log(\tanh\rho)$ and the original 
$H_{\mbox{\scriptsize SS}}$ function $\sech 2\rho$ in the function $H$ have 
the same $2e^{-2\rho}$ asymptotic behaviour as $\rho\rightarrow\infty$. 
Consequently, the $\rho\rightarrow\infty$ asymptotic form of the eigenfunction 
problem remains unchanged with respect to the undeformed Salam-Sezgin 
system, except that the edge of the continuous spectrum is shifted to 
$\lambda=1+k$.}  as we saw for $\Psi_\lambda$ in \eqref{fallingonly}, 
normalizability considerations with respect to the norm 
\eqref{fullnorm} as $\rho\to0$ do not fix the value of $\varpi$ for the 
asymptotic limit of the transverse $\xi(\rho)$ wavefunction. This is due 
to the $\sinh2\rho$ factor in $\sqrt{g_{\rm\sst{EH}}}$ for the 
Eguchi-Hanson metric \eqref{ehmet}, which allows any value of 
$\varpi$ to correspond to a normalizable wavefunction.
The boundary condition for $\varpi$ has to come instead from a careful 
consideration of the effects of the delta-function source term 
in \eqref{sourcedeigenvalue}.

Even taking into account the sourced equation \eqref{sourcedeigenvalue}, 
determining the asymptotic value of $\varpi$ is somewhat elusive. The 
obvious thing to try to do is to multiply \eqref{sourcedeigenvalue} 
by $\sqrt{g_{\rm\sst{EH}}}$ and then to integrate over a small volume 
surrounding $\rho=0$ (this volume corresponds to a $D_2$ disk in 
$(\rho,\chi)$ times an $S^2$ sphere for the angular directions 
$(\theta,\varphi)$). However, noting that $H$ has the asymptotic 
behaviour \eqref{Hasymp} as $\rho\to0$, one finds that the 
$b_\lambda\log\rho$ asymptotic part of \eqref{xiasymptotics} simply 
reproduces the $k$ to $T$ relation \eqref{kTrel} while the $a_\lambda$ 
part of \eqref{xiasymptotics} is eliminated in the $\rho\to0$ limit 
after division by $H$ on the right-hand side, and similarly drops out of 
the left-hand side of the integral of \eqref{sourcedeigenvalue}.

In order to determine the asymptotic value of $\varpi$ for the transverse 
wavefunction, one needs to be more careful and employ a regularised 
approach to the $\delta^2(\rho,\chi)$ delta function in the sourced 
equation \eqref{sourcedeigenvalue}.  The support 
domain of the delta-function source needs to be expanded out slightly, 
to become a ring at radius $\rho=\epsilon$.  Subsequently, we will 
take a limit in which $\epsilon$ is shrunk to zero. Thus, one replaces 
the pointlike delta-function at the center of the $D_2$ integration 
disk by a ring delta-function 
\be
d^2z\delta^2(z)=\frac1{2\pi}d\rho d\chi 
\delta(\rho-\epsilon)\,.\label{ringdelta}
\ee
Integrating the NS5-sourced equation \eqref{sourcedeigenvalue} over a 
small volume including the delta-function source extending from 
$\epsilon_-<\epsilon$ to $\epsilon_+>\epsilon$, one gets
\be
\int_{\epsilon_-}^{\epsilon_+}\sqrt{g_{\rm\sst{EH}}}\,
\triangle_{\rm\sst{EH}}\xi_\lambda = 
-16\pi G_{\sst{10}}Tg^2\int_{\epsilon_-}^{\epsilon_+}
d\rho{\xi_\lambda\over H}\delta(\rho-\epsilon)\,.\label{flucteqn}
\ee
Performing the integral and recalling that 
$H\to-k\log\rho$ as $\rho\to0$, one gets 
\be
\frac\pi2\rho{\partial\over\partial\rho}\xi_\lambda
\Big|_{\epsilon_-}^{\epsilon_+} = {16\pi G_{\sst{10}}Tg^2\over k}
\int_{\epsilon_-}^{\epsilon_+}{\xi_\lambda\over\log\rho}
\delta(\rho-\epsilon)d\rho\,. \label{eggfluctrel}
\ee

Note that inside the ring delta-function source, the wavefunction 
solution $\xi_\lambda$ must be singularity-free, and thus must be 
entirely composed of the constant asymptotic solution 
\be
\xi_\lambda(\rho)\vert_{\hbox{\small in}}=a_\lambda\,,
\ee
so that inside the source one has ${\partial\xi_\lambda\over\partial\rho}\vert_{\hbox{\small in}}=0$.
Letting the asymptotic value of the $\xi_\lambda(\rho)$ wavefunction 
outside the source be
\be
\xi_\lambda(\rho)\vert_{\hbox{\small out}}=
\tilde a_\lambda+b_\lambda\log\rho\,,
\ee
one accordingly has the continuity relation for the undifferentiated 
wavefunction
\be
\xi_\lambda(\rho)\vert_{\hbox{\small in}}=a_\lambda=
\tilde a_\lambda+b_\lambda\log\epsilon=
\xi_\lambda(\rho)\vert_{\hbox{\small out}}\,,
\ee
while from \eqref{eggfluctrel} one obtains the discontinuity in 
${\partial\xi_\lambda\over\partial\rho}$:
\be
\rho{\partial\xi_\lambda\over\partial\rho}\Big\vert_{\epsilon_-}^{\epsilon_+}
=b_\lambda={32G_{\sst{10}}Tg^2\over k}{(\tilde a_\lambda+
  b_\lambda\log\epsilon)\over\log\epsilon}\,,
\ee
which implies
\be
b_\lambda\left(1-{32G_{\sst{10}}Tg^2\over k}\right)=
{32G_{\sst{10}}Tg^2\over k}{\tilde a_\lambda\over\log\epsilon}\,.
\ee
Hence the regularised relation between the $\rho\to0$ asymptotic 
coefficients in the solution outside the source is
\be
\tilde a_\lambda=b_\lambda({k\over32G_{\sst{10}}Tg^2}-1)
\log\epsilon\,.\label{regabrel}
\ee

At the same time, the relation \eqref{kTrel} between $k$ and $T$ is now 
also modified. Instead of relation \eqref{staticrel}, one now has
\be
\int_{\epsilon_-}^{\epsilon_+}d\rho 
\sqrt{g_{\rm\sst{EH}}}\,\triangle_{\rm\sst{EH}}\tilde H=
 -16\pi G_{\sst{10}}Tg^2\int_{\epsilon_-}^{\epsilon_+}
  d\rho\delta(\rho-\epsilon)=-16\pi G_{\sst{10}}Tg^2\,.
\ee
Recalling that inside the source the solution has to be singularity-free, 
so $\tilde H_{\hbox{\small in}}=0$, \ie inside the source there is 
no NS-5 brane back-reaction, while outside the source one has the fully 
back-reacted solution with $H=\tilde H + H_{SS}$, one can perform the 
 integral on the left-hand side to obtain
\be
\frac\pi2\rho{\partial\over\partial\rho}(-k\log\tanh\rho)
\Big\vert_{\epsilon_-}^{\epsilon_+}=-16\pi G_{\sst{10}}Tg^2\,.
\ee
Now
\be
\rho{\partial\over\partial\rho}(-k\log\tanh\rho)={
\rho\over\tanh\rho}\left(1-\tanh^2\rho\right)\rightarrow 1-\frac23\rho^2
\ee
as $\rho\to0$, so the regularised $k\leftrightarrow T$ relation becomes
\be
-k(1-\frac23\rho^2)=-32G_{\sst{10}}Tg^2\,,
\ee
which implies
\be
{k\over 32G_{\sst{10}}Tg^2}-1=\frac23\epsilon^2 + 
   {\cal O}(\epsilon^4)\,.\label{regulatedkTrel}
\ee

If one takes the limit $\epsilon\to0$ at this point in 
\eqref{regulatedkTrel}, one simply reobtains \eqref{kTrel}. However, 
a careful derivation of the boundary condition for $\varpi$ from the 
sourced fluctuation problem (\ref{sourcedeigenvalue},\,\ref{ringdelta}) 
requires combining the regularised $k\leftrightarrow T$ relation with 
the regularised exterior $a_\lambda \leftrightarrow b_\lambda$ 
relation \eqref{regabrel}. Putting these two equations together, one 
obtains
\be
a_\lambda=\frac23 b_\lambda\epsilon^2\log\epsilon + 
{\cal O}(\epsilon^4)\,,\label{regulatedatob}
\ee
and so, upon finally taking the limit $\epsilon\to0$, 
one finds the requirement 
\be
a_\lambda/b_\lambda\rightarrow 0\  \  \mbox{\ie}\  
\varpi(\lambda)=0\,,\label{varpilimit}
\ee
which for bound states corresponds uniquely to the eigenvalue $\lambda=0$. 

Hence the unique bound-state wavefunction for 4D gravitational fluctuations in the presence of the NS5-brane is the zero-mode 
\be
\xi_0(\rho)=\log(\tanh\rho)\,,\label{xizeroagain}
\ee
which, agreeably, is exactly the same as that found in \eqref{zeromode} 
in our preliminary Salam-Sezgin background analysis, prior to the inclusion 
of the NS5-brane.
This is a key result of this paper: recalling that $m^2=g^2\lambda$, we 
have found that a careful treatment of the NS5-brane source for the 
transverse part of the gravitational fluctuation wavefunction shows that 
linearised 4D fluctuations consistent with the conditions imposed by the 
NS5-brane source are {\em massless} in the 4D subspace.

\subsection{Asymptotic conformal invariance and 
self-adjointness}\label{ssec:selfadjoint}

The delicacy needed in analysing the $\rho\to0$ boundary condition 
\eqref{varpilimit} reflects the specific asymptotic character of the radial 
Schr\"odinger problem 
(\ref{schrodinger},\,\ref{SSpotential}), both in the original undeformed 
Salam-Sezgin background and also after inclusion of the NS5-brane. Taking 
the $\rho\to0$ limit of the potential \eqref{SSpotential}, one obtains
\be
V_{\sst \rho\to0}=-{1\over4\rho^2}\,.\label{asympV}
\ee
The corresponding one-dimensional quantum mechanical problem has a long 
history \cite{Case:1950an,Landau1960,de Alfaro:1976je}. A good overview is 
given in \cite{Essin2005}. The special character of this one-dimensional 
problem involves not only the $1/\rho^2$ form of the potential, which 
gives rise to an ${\rm O}(2,1)$ 1D conformal invariance, but also 
the $-\frac14$ coefficient, which is a critical value. For a 
potential $V=\gamma/\rho^2$ with coefficient $\gamma>-\frac14$, a 
regularised treatment shows that there is no $L^2$ normalizable bound 
state, while for $\gamma<-\frac14$, an infinity of discrete $L^2$ 
normalizable bound states appears. At the critical value 
$\gamma=-\frac14$, there is just a single bound state.

The character of the $V=-1/(4\rho^2)$ Schr\"odinger problem is exactly 
as we have found above in Section \ref{ssec:fluctuations}: the 
requirement of normalisation does not fix the asymptotic form of a 
bound-state wavefunction at the origin. A candidate 
bound-state wavefunction for a general value of $\lambda$ would 
spontaneously break the asymptotic 1D conformal invariance. The one 
exception to this is the $\lambda=0$ wavefunction that we found in 
\eqref{varpilimit}.

Another key feature of the $V=-1/(4\rho^2)$ Schr\"odinger problem is the 
delicate issue of self-adjointness of the corresponding Hamiltonian, or, 
in our case, of the $-{d^2\over d\rho^2}$ part of the wave operator 
\eqref{schrodinger}. For two normalizable candidate bound-state 
wavefunctions $\Psi_1$ and $\Psi_2$, self-adjointness of this operator 
requires $\int_0^\infty (\Psi_1^\ast{d^2\over d\rho^2}\Psi_2 
-\Psi_2 {d^2\over d\rho^2}\Psi_1^\ast) =0$, which in turn requires 
\be
(\Psi_1^\ast{d\over d\rho}\Psi_2 - 
  \Psi_2 {d\over d\rho}\Psi_1^\ast){\Big|}_0^\infty=0\,.\label{selfadjcond}
\ee
For normalizable bound-state wavefunctions, there is no problem with this 
requirement as $\rho\to\infty$, but for $\Psi_1$ and $\Psi_2$ having 
$\rho\to0$ asymptotic structure $\sqrt\rho(a_i+b_i\log\rho)$ 
(corresponding to \eqref{xiasymptotics} for rescaled wavefunctions 
\eqref{rescaledwavefunction}, with $i=1,2$) 
the condition \eqref{selfadjcond} requires
\be
\frac{a_1}{b_2}=\frac{a_2}{b_2}\,,\label{selfadjcoeffcond}
\ee
\ie $\Psi_1$ and $\Psi_2$ must have the same eigenvalue $\lambda$, 
since $a_\lambda/b_\lambda$ is a single-valued function of $\lambda$, as 
we have seen.\footnote{For scattering-state wavefunctions, condition 
\eqref{selfadjcond} and hence \eqref{selfadjcoeffcond}  are also required, 
but the requirements as $\rho\to\infty$ are different for delta-function 
normalizable states and no single scattering-state eigenvalue 
$\lambda$ is selected.}

This is the underlying reason for the existence of precisely one 
bound-state eigenfunction. In our case, coupling to the NS5-brane source 
as in \eqref{sourcedeigenvalue} selects $\lambda=0$. In the case of the 
classic $V=-1/(4\rho^2)$ Schr\"odinger problem, this would not be very 
good, because it would put the single allowed bound state right at the edge 
of the continuum of scattering states. However, in the present case, 
the potential \eqref{SSpotential} deviates from the $-1/(4\rho^2)$ 
structure as $\rho$ increases away from zero. This has the effect of 
raising the edge of the continuous spectrum up to $\lambda=1+k$, as we have 
seen above in Sections \ref{ssec:schrodingereqn} and \ref{ssec:fluctuations}.

\section{The braneworld Newton constant}\label{sec:Newtonconstant}

Having established that there exists a zero-eigenvalue bound-state 
transverse wave function \eqref{zeromode}, the basic aim of constructing 
a type IIA supergravity brane configuration that localises massless gravity 
in a four-dimensional brane subspace has been achieved. A key achievement 
of this construction is the non-zero value of Newton's constant for the 
massless effective gravity theory in the 4D subspace, despite the
infinite volume of the transverse space. Starting from the Einstein-frame 
gravitational action for the 10D metric $\hat g_{MN}$
\be
I_{\sst{10}}={1\over 16\pi G_{\sst{10}}}\int d^{10}x \sqrt{\hat g} 
\hat R(\hat g)\,, \label{10act}
\ee
the effective theory for 4D gravitational fluctuations is obtained 
starting from the Einstein-frame form $ds_{10\,\rm ein}^2=
e^{-\phi/2}ds_{10\,\rm str}^2$ of the static string-frame metric given 
in \eqref{ns5brane} and making the replacement 
$\eta_{\mu\nu}\rightarrow\eta_{\mu\nu}+h_{\mu\nu}(x)\xi(\rho)$, as we 
have done above in Sections  \ref{sec:boundstates} and 
\ref{sec:sourceinclusion}. The angular coordinates $\psi,\theta,\phi$ and 
$\chi$, for which we are considering only S-waves without further 
excitation, give rise to compact integrals producing corresponding 
factors of $g$ (with length dimension $-1$) in the effective theory. 
In order to obtain a 4D effective theory, we compactly also the $y$ 
coordinate with a circumference $\ell_y$. Recalling that $\psi$ takes 
values in the range $[0,4\pi)$, one finds at quadratic order in 
$h_{\mu\nu}(x)$ an effective action for 4D linearised gravity
\be
I_{\rm\sst{lin\,4}}={1\over\upsilon_0^2}\int d^4x 
\left(-\frac12\partial_\sigma h_{\mu\nu}
\partial^\sigma h^{\mu\nu}+\frac12\partial_\mu 
h^\sigma{}_\sigma\partial^\mu h^\tau{}_\tau + 
\partial^\nu h_{\mu\nu}\partial^\sigma h^\mu{}_\sigma + 
h^\sigma{}_\sigma\partial^\mu\partial^\nu h_{\mu\nu}\right)\,,\label{lin4D}
\ee
where
\be
\upsilon_0=
\left({16\pi G_{\sst{10}}g^5\over\pi^2\ell_yI_2}\right)^{\frac12}\,,
\label{upsilon0}
\ee
in which
\be
I_2=\int_0^\infty d\rho H\sinh2\rho\cosh2\rho\,\xi_0^2=
\frac{\pi^2}{24}(2+3k)\,.\label{I2}
\ee
Note that, up to a factor ${2\pi^3\over g^5}$, $I_{\sst{2}}$ is just 
the $(\hbox{norm})^2$ \eqref{fullnorm} of $\xi_0$. Combining 
Eqns \eqref{upsilon0} and\eqref{I2}, one has the normalisation factor
\be
\upsilon_0=\left({384G_{\sst{10}}g^5\over\pi^3\ell_y(2+3k)}
    \right)^{\frac12}\,.\label{normedupsilonzzero}
\ee

If it were not for the normalizable character of the zero-mode 
$\xi_0(\rho)$, one would obtain a vanishing value for $\upsilon_0$. This 
is what would happen for a standard Kaluza-Klein reduction to the 
$\rho$-independent sector of the theory, as in \cite{Cvetic:2003xr}, where
$\xi_0(\rho)$ is simply a constant. 
In order to calculate the effective 4D Newton's constant, one now needs to 
rescale $h_{\mu\nu}=\upsilon_0\tilde h_{\mu\nu}$ in order to obtain a 
canonically-normalised kinetic term \eqref{lin4D} for $\tilde h_{\mu\nu}$. 
Then the leading effective 4D coupling $\kappa_4=\sqrt{32\pi G_{\sst4}}$ 
for gravitational self-interactions is obtained from the coefficient in the
trilinear 
terms in $\tilde h_{\mu\nu}$ in the 4D effective action.
These involve an integral\footnote{The specific form \eqref{xizeroagain} 
of the $\xi_0$ zero-mode has the agreeable property that the coefficients 
of yet higher-order terms in $h_{\mu\nu}(x)$ in the 4D effective action 
can also be explicitly evaluated. One finds $I_{p} \equiv 
\int_0^\infty d\rho\, H\, \sinh2\rho\, \cosh 2\rho\, \xi_0^p = 
(-1)^p\, p!\,2^{-p-1}\, \zeta(p)\, [2 + (p+1)\, k]$.}
\be
I_3=\int_0^\infty d\rho H\sinh2\rho\cosh2\rho\,\xi_0^3=
-\fft{3(1+2k)\, \zeta(3)}{4}\,.\label{I3}
\ee
The $\sinh2\rho$ factor in \eqref{I3}, arising from $\sqrt{g_{\rm\sst{EH}}}$, 
leads to the convergence of \eqref{I3} in the limit as $\rho\to0$, just as 
it does for $I_2$. The 4D massless gravitational coupling 
$\kappa_4^{(0)}=\sqrt{32\pi G_{\sst4}}$ is then obtained upon 
rescaling $h_{\mu\nu}=\upsilon_0\tilde h_{\mu\nu}$ so as to obtain a 
conventionally normalised quadratic action for $\tilde h_{\mu\nu}$, and 
then extracting the coefficient of the trilinear $\tilde h_{\mu\nu}$ terms 
in the effective action:
\be
\kappa_4^{(0)}=\left|\left({16\pi G_{\sst{10}}
g^5\over\pi^2\ell_y}\right)^{\frac12}{I_3\over(I_2)^{\frac32}}\right|
\,.\label{kappa4integrals}
\ee
Using \eqref{I2} and \eqref{I3}, this becomes
\be
\kappa_4^{(0)}=144\sqrt6\zeta(3)\left({G_{\sst{10}}g^5\over\pi^7\ell_y}
\right)^{\frac12}{(1+2k)\over(2+3k)^\frac32}\,,\label{kappa4}
\ee
and so the 4D Newton constant is given by
\be
G_{\sst4}={3888\,\zeta(3)^2G_{\sst{10}}g^5\over\pi^8\ell_y}{(1+2k)^2
\over(2+3k)^3}\,.
\ee

Evaluating gravitational couplings to matter, as opposed to the 
gravitational self-coupling, requires setting up a model of 
non-gravitational matter in the 4D subspace. One approach to this would be 
to employ a Ho\v{r}ava-Witten construction \cite{Horava:1996ma}, replacing 
the $S^1$ compactification of the $y$ direction by a $S^1/\Z_2$ orbifold. 
One way to view this would be as a 7D to 6D reduction starting from the 
7D theory on ${\cal H}^{(2,2)}$ \cite{Cvetic:2003xr}, thus obtaining a 
6D chiral $(1,0)$ supersymmetric theory with potential anomalies arising 
from anomaly inflow, and hence requiring compensating 6D matter 
fields \cite{Pugh:2010ii}. Another way to view it would be as a 5D to 4D 
reduction after the additional $S^2\ \&\ \hbox{monopole}$ reduction from 
7D, thus producing a 4D chiral N=1 supersymmetric theory, again with 
anomaly inflow requiring compensating 4D matter fields \cite{Lukas:1999nh}. 
Either of these two $S^1/\Z_2$ approaches would have the additional effect 
of reducing the final surviving supersymmetry to $N=1$ in 4D, which 
could be of practical physical interest. A simpler approach to modelling 
matter fields, however, which is all that we shall consider here, is just 
to consider the non-gravitational 4D fields accompanying gravity in the 
descent to 4D from 10D type IIA supergravity. From the string-frame 
unbroken supersymmetry \eqref{kspin2}, one sees that superpartners of the 
graviton should involve the same $\xi_0$ transverse wavefunction, giving 
rise to bilinear kinetic and trilinear gravitational effective-action terms 
involving the same $I_2$ and $I_3$ integrals as for the gravitational 
self-coupling, and hence the same $\kappa_4$ gravitational coupling 
constant \eqref{kappa4}.

\section{Corrections to 4D Newtonian gravity}\label{sec:gravitycorrections}

Finally, let us sketch the consequences of the continuum of $\lambda\ne0$ 
transverse gravitational eigenmodes. The corresponding 4D 
$h_{\mu\nu}^{(\lambda)}$ massive gravitational modes are separated from 
the massless 4D gravitational states with transverse eigenmode $\xi_0$ 
by a gap in $(\hbox{mass})^2$ eigenvalues of height $(1+k)g^2$, as we 
have seen. For $\lambda>1+k$, as seen in Section \ref{ssec:schrodingereqn} 
(but now with the NS5-brane moving the edge of the continuum from 
$\lambda=1$ to $\lambda=1+k$), the transverse $\xi_\lambda(\rho)$ 
eigenfunctions have oscillatory behaviour as $\rho\to\infty$, instead of 
the rising or falling exponential behaviour of the candidate bound states 
for $\lambda<1+k$. The boundary-condition implications of the NS5-brane 
source as analysed in Section \ref{ssec:fluctuations} remain valid also 
for $\lambda\ne0$ eigenmodes: a general Frobenius analysis shows that 
their $\rho\to0$ asymptotics have to be as in \eqref{xiasymptotics}, but the 
constraints of the NS5-brane source imply that the asymptotic constant 
part of a $\xi_{\lambda\ne0}$ wavefunction must vanish, just as it must 
for $\xi_0$. For the candidate $\lambda<1+k$ bound states, it is only 
for $\lambda=0$ that this boundary condition proves to be consistent 
with the other boundary condition needing to be imposed as 
$\rho\to\infty$: elimination of the most weakly falling exponential term, 
in order to obtain normalizability. However, for the continuum of 
$\lambda>1+k$ wave functions, one does not demand standard normalizability 
with respect to the norm \eqref{fullnorm}. Instead, just as in 
free-field theory, continuum wavefunctions need to be delta-function 
orthonormalised. 

Gravitational fluctuations involving $\lambda\ne0$ transverse 
$\xi_\lambda$ eigenmodes make small contributions to the 4D effective action. 
Starting from the edge of the continuous transverse spectrum at 
$\lambda=1+k$, one sees from \eqref{eigenvalueproblem} that the spectrum 
of gravitational modes arising from the transverse dynamics will have 
continuous $(\hbox{mass})^2$ eigenvalues ranging over the interval 
$(1+k)g^2 \le m^2 \le \infty$. Repeating the normalisation analysis 
for such $\xi_\lambda$ modes, one finds that in order to have canonically 
normalised kinetic terms, the $h_{\mu\nu}^{(\lambda)}$ massive graviton 
fields require rescaling by
\be
\upsilon_\lambda=\left({16\pi G_{\sst{10}}g^5\over\pi^2\ell_y}
\right)^{\frac12}{\cal N}_\lambda\,,\label{upsilonlambda}
\ee
where ${\cal N}_\lambda$ is a normalisation coefficient depending on 
the details of $\xi_\lambda$.

Assuming that matter interacting with the continuum of massive 
gravitational modes itself has transverse wavefunction $\xi_0$, its 
interaction with such massive gravitational modes at the trilinear level 
involves  an integral
\be
I_{2,\lambda}=\int_0^\infty d\rho H\sinh2\rho\cosh2\rho\, \xi_0^2\,
\xi_\lambda\,,
\ee
which is convergent as $\rho\to0$ owing to the $\sinh2\rho$ term, 
and as $\rho\to\infty$ owing to the $e^{-2\rho}$ asymptotic falloff 
of $\xi_0$. Rescaling all fields in order to produce canonical kinetic 
terms then gives rise to the coupling between $h_{\mu\nu}^{(\lambda)}$ 
and 4D matter:
\be
\kappa_4^{(\lambda)}=\left|\left({16\pi G_{\sst{10}}g^5\over\pi^2\ell_y}
\right)^{\frac12}{I_{2,\lambda}\over I_2}{\cal N}_\lambda\right|=
{24\over\pi^2(2+3k)}
\left|\left({16\pi G_{\sst{10}}g^5\over\pi^2\ell_y}
\right)^{\frac12}I_{2,\lambda}\,{\cal N}_\lambda\right|\,.
\ee

Any given continuum massive gravitational mode will produce a 
Yukawa correction
\be
\Delta V_\lambda = -(\kappa_4^{(\lambda)})^2M_1M_2{
       e^{-g\sqrt\lambda |x|}\over|x|}
\ee 
to the 4D Newtonian $V=-(\kappa_4^{(0)})^2M_1M_2/|x|$ potential, 
where $|x|$ is the distance between masses $M_1$ and $M_2$ in the 4D 
subspace. If one assumes that $\kappa_4^{(\lambda)}$ does not have a 
strong dependence on $\lambda$ for $\lambda\gtrapprox1+k$, and noting 
that for large $\lambda$ the falling exponential suppresses 
$\Delta V_\lambda$ contributions, then one obtains an integrated 
correction to the Newtonian potential 
\bea
\Delta V &=& \int_{1+k}d\lambda \Delta V_\lambda\nn\\
&\simeq& - {2M_1M_2(\kappa_4^{(1+k)})^2e^{-g\sqrt{(1+k)}|x|}\over g|x|^2}
\left((1+k)^{\frac12} + {1\over g|x|}\right)\,.
\eea
Note that the corrections to the Newtonian potential have leading 
behaviour ${e^{(-g\sqrt{(1+k)}|x|)}\over g|x|^2}$ here instead of the ${1\over |x|^3}$ leading correction in \cite{Randall:1999vf},
because the edge of the continuous spectrum in the present construction is 
located at $m^2=g^2(1+k)$ instead of $m^2=0$.

\section{Conclusion}\label{sec:conclusion}

The hyperbolic background with NS5-brane construction that we have given 
in this paper provides a successful localisation of gravity in a 
lower-dimensional subspace of type IIA supergravity. In doing so, it has 
evaded a number of problems with braneworld gravity localisation that have 
been raised in the literature. A fuller analysis will be needed to understand 
how this construction may be generalised to other situations, but we may 
identify a number of features of our construction that help to evade 
some of the problems that have been raised. In the discussions that arose 
following Ref.\ \cite{Randall:1999vf}, which was based on the junction of 
AdS slices, a ``no-go'' theorem was put forward in 
Ref.\ \cite{Maldacena:2000mw}. This ruled out braneworld reductions that 
are non-singular or with singularities satisfying an admissibility 
criterion that the $g_{00}$ component of the metric should not increase 
as one approaches the singularity (which our NS5-brane construction 
satisfies). However, that analysis assumed that the scalar-field potential 
is non-positive. This is clearly not the case for the ${\rm SO}(2,2)$ 
invariant potential of the 7D theory \cite{Cvetic:2003xr} obtained 
{\it via} ${\cal H}^{(2,2)}$ reduction, which is positive-definite, 
generalising the positive potential of the Salam-Sezgin model 
\cite{Salam:1984cj}. The positivity of this potential is key to allowing 
a solution incorporating an $S^2$ and 4D flat space -- otherwise, reduction 
on an $S^2$ factor would give rise to an anti-de Sitter space in 4D.
The noncompact ${\rm SO}(2,2)$ invariant structure is a consequence of 
the underlying hyperbolic geometry, which thus appears to be a key feature 
of our successful construction.

A more substantial potential problem with braneworld localisations related 
to our construction was outlined in Ref.\ \cite{Bachas:2011xa}. Indeed, 
a simple argument was given there that would seem to imply that a 
localisation producing massless effective gravity in the lower dimension of 
a spacetime with infinite transverse geometry can only be made with a 
constant transverse wavefunction. This would seem to rule out normalisable 
bound states yielding massless gravity. However, a key feature of that 
argument involves integration by parts of one of the derivatives in the 
transverse wave operator. In the corresponding integration by parts in 
our construction, one cannot ignore the surface term, and so the 
demonstration that the wavefunction $\xi_0$ has to be constant fails in 
our case. This does, however raise the issue of self-adjointness of the 
transverse wave operator, which we have discussed in Section 
\ref{ssec:selfadjoint}. Provided there is just {\em one} bound state, 
the wave operator can be self-adjoint, as in our case with the single 
transverse bound state $\xi_0$. So another important feature of our 
construction appears to be this quite special behaviour of the transverse 
eigenvalue problem near the $\rho=0$ ``waist'' of the ${\cal H}^{(2,2)}$ 
space, which also can be viewed as the apex of the underlying 
Eguchi-Hanson space.

The present construction has been purely classical, focusing only on a 
supergravity realisation. Since the only elements used, \ie type IIA 
supergravity and an NS5-brane, are also native objects in string theory, 
the construction carries over directly into string theory. From that point 
of view, the solution for $e^\phi$ in the fully NS5 back-reacted 
solution \eqref{ns5brane} with $H$ given by \eqref{Hsol2}, \ie with an 
asymptotic $e^{-2\rho}$ behaviour, shows that the solution has 
asymptotically $\phi\propto-\rho$: it asymptotically tends to a 
linear-dilaton string-theory vacuum. The eight-supercharge unbroken 
supersymmetry (\ref{kspin},\,\ref{proj}) of the back-reacted solution 
guarantees stability. This remains so should one decide to sacrifice half 
of this supersymmetry either by reducing the $y$ coordinate on 
$S^1/\Z_2$, or by making a corresponding Ho\v{r}ava-Witten construction, 
leaving just $N=1$ supersymmetry in four dimensions. At the quantum level, 
some usual additional considerations will come into play. For example, 
the NS5-brane charge $T$ will have to take its value on a quantised 
charge lattice, following from Dirac-Schwinger-Zwanziger quantisation 
conditions \cite{Nepomechie:1984wu, Teitelboim:1985ya, Bremer:1997qb}.

Duality symmetries will yield other realisations of the construction 
given in this paper. For example, a T-duality transformation will change 
the present 10D type IIA solution into a solution of type IIB supergravity. 
Depending on whether this is done in a worldvolume or in a transverse 
direction (\eg in one of the smeared transverse angle coordinates), 
one can get a 4-brane or an ALF space \cite{Tong:2002rq}. A general 
understanding of the duality maps of the present construction may help 
to categorise situations in which localised braneworld gravity can exist.

The gravity-localising construction given in this paper is not a 
dimensional reduction: at all stages, we have been working with solutions 
of 10D supergravity. Although reduction on ${\cal H}^{(2,2)}$ produces a 
consistent Kaluza-Klein truncation \cite{Cvetic:2003xr} to the 
${\rm SO}(2,2)$ invariant theory containing the Salam-Sezgin model, in 
the present paper we have not sought to eliminate dependence on all the 
transverse coordinates. For simplicity, we have considered ``S-wave'' 
solutions independent of the transverse-coordinate angles 
$\theta,\,\phi$ and $\chi$, but this restriction could straightforwardly 
be relaxed. Our construction depends in a fundamental way, however, on the 
transverse radial coordinate $\rho$, and the effective theory derived from 
the transverse zero-mode $\xi_0(\rho)$ is {\em not} a consistent 
Kaluza-Klein truncation. The continuum of massive modes lying above 
the $g\,\sqrt{1+k}$ mass gap produce small corrections to the 
leading-order massless braneworld gravity. In Section 
\ref{sec:gravitycorrections}, we have made a preliminary sketch of these 
corrections, but this question clearly deserves a more complete study. 

\section*{Acknowledgements} 
 
 We would like to thank Costas Bachas, Guillaume Bossard, John Estes, 
Hong L\"u and Dan Waldram for helpful discussions. For hospitality over the course of the 
work, K.S.S.\ and C.N.P.\ would like to thank
the Cynthia and George Mitchell Foundation and the Cambridge--Mitchell collaboration for a workshop at 
Great Brampton House, Madley, Herefordshire, and also the KITPC, Beijing.  
K.S.S. would also like to thank the Mitchell Institute, 
Texas A\&M University, College Station, the Perimeter Institute, 
Waterloo, and the Albert Einstein Institute, Potsdam. 
The work of K.S.S.\ was supported in part by the STFC under Consolidated 
Grant ST/J0003533/1; the work of C.N.P.\ was supported in part by 
DOE grant DE-FG02-13ER42020 and that of B.C.\ was supported by an 
STFC PhD studentship.

\end{document}